\documentclass[final, envcountsect]{svjour3}

\usepackage{ae, lmodern}					
\usepackage[english]{babel}
\usepackage[utf8]{inputenc}
\usepackage[T1]{fontenc}

\usepackage[top=3cm, bottom=3cm, left=3cm, right=3cm]{geometry}

\usepackage{amsfonts, amsmath, amssymb}
\numberwithin{equation}{section}

\usepackage[hidelinks, pdfauthor=author]{hyperref}

\usepackage{tabularx, enumitem, verbatim, xcolor, epsfig, float, cleveref}
\usepackage{marvosym}

\Crefformat{figure}{#2Fig.~#1#3}
\Crefmultiformat{figure}{Figs.~#2#1#3}{ and~#2#1#3}{, #2#1#3}{ and~#2#1#3}

\usepackage{pgfplots, tikz, tikzscale}
\pgfplotsset{compat=1.14}

\newcommand{\subf}[2]{{\small \begin{tabular}[t]{@{}c@{}} #1\\#2 \end{tabular}}}

\newcommand{\R}{{\mathbb R}}
\newcommand\ten[2][1]{
  	\ifnum#1=2 {\boldsymbol{#2}} \fi%
  	\ifnum#1=3 {\textit{\textbf{#2}}} \fi%
  	\ifnum#1=4 {\mathbb #2} \fi}

\newcommand{\norm}[1]{\left \| #1 \right \|}
\newcommand{\dpair}[2]{\left \langle #1 \mid #2 \right \rangle}
\newcommand{\mean}[1]{\left \langle#1 \right \rangle} 
\newcommand{\newboxsymbol}[2]{
\begin{tikzpicture}
\filldraw[fill=#1,draw=#2] circle (4pt);
\end{tikzpicture}}

\definecolor{crazyblue}{RGB}{37,84,138}
\definecolor{crazyyellow}{RGB}{255,243,191}

\journalname{Continuum Mechanics and Thermodynamics}

\title{Design and testing of 3D printed micro-architectured polymer materials exhibiting a negative Poisson's ratio}
\titlerunning{Design and testing of 3D printed micro-architectured polymer materials} 

\author{Filippo Agnelli \and Andrei Constantinescu \and Grigor Nika}
\authorrunning{Agnelli et al.}

\institute{
Filippo Agnelli \and Andrei Constantinescu (\Letter) \at
\textsc{LMS, CNRS, \'Ecole polytechnique, Institut polytechnique de Paris, 91128 Palaiseau, France.}
\\
\email{andrei.constantinescu@polytechnique.edu}
\and
Grigor Nika \at
\textsc{Weierstrass Institute for Applied Analysis and Stochastics, 10117 Berlin, Germany}
}

\date{Received: 19 July 2019 / Accepted: 13 November 2019 / Published online: 20 November 2019}

\def\makeheadbox{{%
\hbox to0pt{\vbox{\baselineskip=10dd\hrule\hbox
to\hsize{\vrule\kern3pt\vbox{\kern3pt
\hbox{\bfseries Continuum Mechanics and Thermodynamics manuscript No. CMAT-D-19-00116}
\hbox{This is a post-peer-review, pre-copyedit version of this article. Version \today.}
\hbox{The final authenticated version is available online at: \href{https://doi.org/10.1007/s00161-019-00851-6}{https://doi.org/10.1007/s00161-019-00851-6}.}\kern3pt}\hfil\kern3pt\vrule}\hrule}%
\hss}}}

\begin{document}

\maketitle

\begin{abstract}
This work proposes the complete design cycle for several auxetic materials where the cycle consists of three steps (i) the design of the micro-architecture, (ii) the manufacturing of the material and (iii) the testing of the material. Topology optimisation via a level set method and asymptotic homogenisation permit to obtain periodic micro-architectured materials with a prescribed effective elasticity tensor and Poisson's ratio. The space of admissible micro-architectural shapes that carries orthotropic material symmetry allows to attain shapes with an effective Poisson's ratio below -1. Moreover, the specimens are manufactured using a commercial stereo-lithography Ember printer and are mechanically tested. The observed displacement and strain fields during tensile testing obtained by digital image correlation match the predictions from the finite element simulations and demonstrate the efficiency of the design cycle.
\end{abstract}


\section{Introduction} \label{intro}
The Poisson’s ratio $(\nu)$ is a measure of the relative amount a given material contracts transversally under a uniaxial stretch loading \cite{Greaves2011}. Unlike most conventional materials, auxetic materials tend to expand transversely to an applied uniaxial stretch load and vice versa, leading to a so-called negative Poisson’s ratio. This effect occurs due to the particular internal microstructure and due to the mechanisms of deformation when loaded. Because of their special mechanical properties, tailored auxetic materials can display enhanced stiffness and energy absorption capabilities \cite{Imbalzano2016,Li2018}, indentation resistance \cite{Lakes1993a}, greater fracture toughness \cite{Choi1996}, crashworthiness \cite{Hou2015}, phononic performance \cite{Spadoni2009} as well as many other interesting properties, making them suitable in targeted applications \cite{Duncan2018,Saxena2016}.\smallskip\\
Since the seminal works performed in the '80s, the design of periodic auxetic structures has attracted research interests and several types of auxetic materials have been introduced. In 1985, Almgren introduced a re-entrant honeycomb structure with Poisson's ratio of -1 using rods, hinges, and springs \cite{Almgren1985}. The re-entrant honeycomb structure was also introduced as a ``bi-mode'' extremal material which supports a stress with a negative determinant in Milton \& Cherkaev \cite{Milton1995}. Conceptual designs of composite materials with Poisson's ratio approaching -1 were presented in \cite{Milton1992}. Some important features of auxetic materials, such as the re-entrant corners, were discussed in the key works of Lakes \cite{Lakes1987,Lakes1993}, Friis et al. \cite{Friis1988} and Evans \cite{Evans1991}. A new class of auxetic materials, that obtain their auxetic behaviour from the \textit{rotating squares} mechanism were introduced in the works of Grima et al. \cite{Grima2007}. Moreover, a separate class of 3D auxetics that exploits the buckling mechanism in structures was proposed in \cite{Babaee2013}.\smallskip\\
Design techniques using modern numerical methods such as shape and topology optimisation \cite{Allaire2002,Bendsoe2004} arose as a natural way to tailor mechanical properties through a design of complex geometries. For materials with a periodic microstructure, the effective elastic properties can be derived by means of asymptotic homogenisation, with periodic boundary conditions applied on a unit cell and the associated boundary value problem resolved \cite{Bakhvalov2011,Mei2010,Sanchez-Palencia1987}. The works of Sigmund presented a topology optimisation framework for designing 2D and stacked 2D auxetic truss-based structures \cite{Sigmund1994} and for a designing 2D continuum-based micro-mechanism with negative Poisson's ratio \cite{Larsen1997}. Since these works, different techniques have been adopted for auxetic structure design optimisation, including SIMP (Solid Isotropic Material with Penalisation), phase-field, level set methods, etc. In the works of Wang et al. \cite{Clausen2015,Wang2014a}, the SIMP method was used to include geometrical non-linearities and to tackle manufacturing constraints. The resulting architectures attain the desired response through uniform features, like the thickness of connecting rods. In more recent developments, the method is extended to thermodynamic topology optimisation or graded materials \cite{Carraturo2019,Jantos2018,Ranaivomiarana2018}. Furthermore, alternative optimisation methods as found in the works of Wang, Mei \& Wang \cite{Wang2004}, Vogiatzis et al. \cite{Vogiatzis2017}, Nika \& Constantinescu \cite{Nika2019}, Wang et al. \cite{Wang2014}, among others, use asymptotic homogenisation, the level set method \cite{Osher1988} and the Hadamard shape derivative to identify material regions and track boundary changes to systematically design auxetic shapes. Wang, Mei \& Wang \cite{Wang2004}, design linear elastic and thermoelastic materials with negative Poisson's ratio, while Nika \& Constantinescu \cite{Nika2019} design linear elastic multi-materials with negative Poisson's ratio.\smallskip\\
The classical theory of elasticity states that isotropic three-dimensional materials may exhibit Poisson's ratios bounded in $[-1,0.5]$ ; Two-dimensional isotropic systems can exhibit Poisson's ratios bounded in $[-1,1]$. The effective elastic tensor that characterises the auxetic material however has a priori orthotropic symmetry. In two-dimensional systems, the effective elastic stiffness is characterised by two Poisson's ratios $\nu_{12}$, $\nu_{21}$, which are a priori not bounded, hence they can assume any positive or negative values in certain directions \cite{Ting2005}. For instance, Poisson's ratios smaller than -1 have been reported according to Lakes \cite{Lakes1993}. In the topology optimisation literature, the auxetic shapes obtained tend to exhibit cubic symmetry, \textit{i.e.} $\nu_{12}$ = $\nu_{21}$. However, as was already mentioned, this need not be the case.\smallskip\\ 
The aim of this paper is to complete a design cycle for several auxetic materials. We combine topology optimisation to systematically obtain the micro-architecture with 3D printing to digitally fabricate the designs and validate against the numerically predicted behaviour. Materials are manufactured using a desktop stereo-lithography 3D printer and then tested on standard tensile machine. Insight into the local mechanical fields is obtained using digital image correlation. The paper is organised as follows. Section 2 presents the computational design of the micro-architectured material. It provides some basic results needed from the theory of homogenisation, relate the effective coefficients to the effective Poisson's ratio, and sets up the optimisation problem to systematically identify optimal auxetic shapes. Section 3 presents the optimal auxetic microstructures obtained and describes some of their properties as well as the additive manufacturing process. Section 4 is dedicated to the experimental testing of the structures and the interpretation of results using Digital Image Correlation (DIC). Additionally, an appendix reviews the approach used to measure the effective Poisson's ratio by DIC and the finite element method on periodic structures.

\paragraph{Notation} Scalars are denoted by italic letters, $\alpha$; Vectors are denoted by bold face letters, $\vec{u}$; Second order tensors are denoted by italic bold face letters, $\ten[2]{\sigma}$; Fourth order tensor are denoted by barred letters, $\mathbb{C}$. The dot product between two second order tensors $\ten[2]{A}$ and $\ten[2]{B}$ is denoted by $\ten[2]{A}:\ten[2]{B} = \sum_{i,j=1}^{N} A_{ij}B_{ji}$ where $A_{ij}$ and $B_{ij}$ are the tensor components. The average of a quantity over a region \textit{e.g.} $D$ is denoted by $\mean{\cdot}_{D}$ while by $\dpair{\cdot}{\cdot}$ we denote the duality product.


\section{Computational design}
The considered micro-architectured materials are two-dimensional periodic assemblies of square unit cells. The unit cells are a two-phase composite with a strong and weak phase, \textit{e.g.} polymer material and void respectively. In the sequel, we will denote as \textit{shape} the strong phase of the micro-architectured unit cell, \textit{e.g.} polymer phase of the composite.\smallskip\\
In this section we present the micro-architectured material modeling using a level set representation \cite{Osher1988} in the unit cell, the computation of effective elastic tensors of the periodic microstructure based on rigourous mathematical theory \cite{Allaire2002,Bakhvalov2011,Mei2010,Sanchez-Palencia1987} and a shape optimisation method \cite{Allaire2004,Wang2004} to reach a target homogenised elasticity tensor.

\subsection{Shape of a unit cell and homogenised elastic tensor}
\begin{figure}
\centering
\begin{tikzpicture}[scale=1.]
\node[right] (phi) at (0.,0.){\includegraphics[scale=0.2]{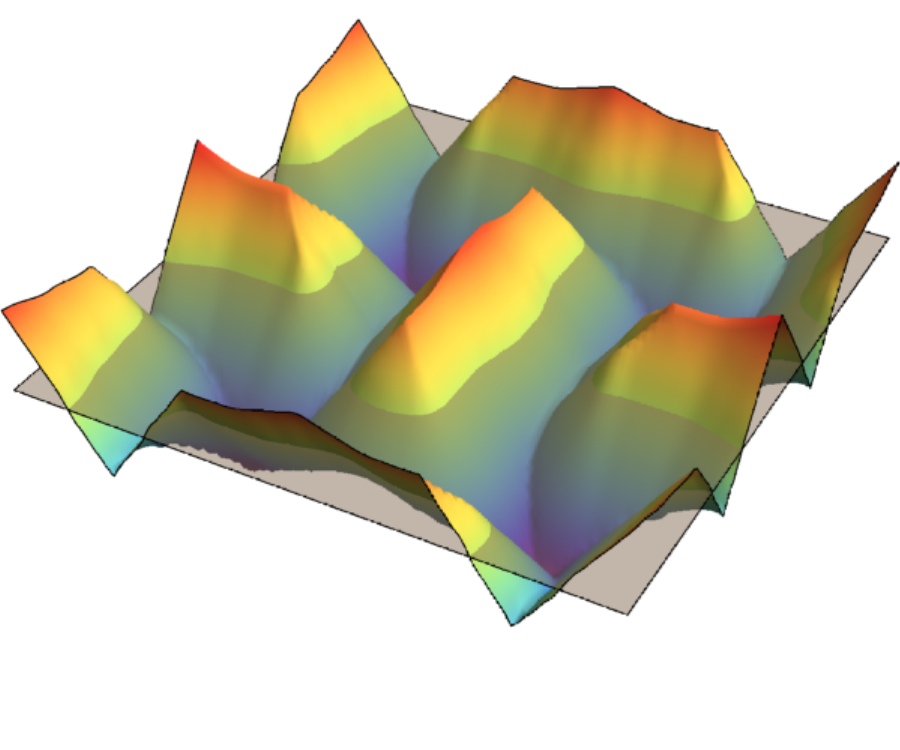}};
\node[right] (domain) at (4.,0.){\includegraphics[scale=0.2]{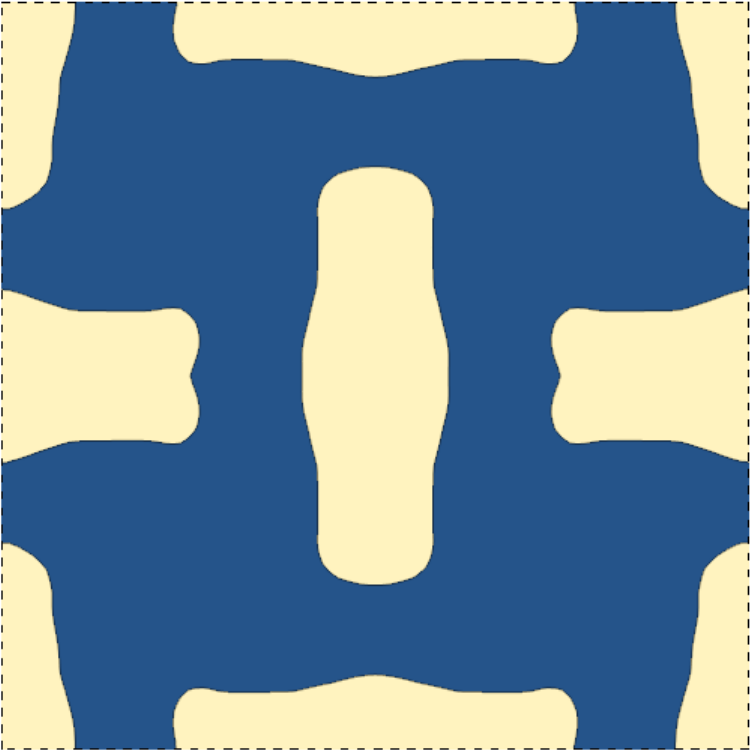}};
\draw[-] (0.,1.0) node [left] {$\phi > 0$} -- (0.8,0.6);
\draw[-] (0.,0.5) node [left] {$\phi = 0$} -- (0.8,0.375);
\draw[-] (0.,0.5)                          -- (1.4,-0.1);
\draw[-] (0.,0.0) node [left] {$\phi < 0$} -- (0.4,-0.2);

\draw (5.1,0.75) node [left,text=white] {\tiny $\phi<0$};
\draw (4.,0.0) node [right] {\tiny $\phi>0$};
\draw[-,color=white] (5.1,-0.75) node [left,text=white] {\tiny $\phi=0$} -- (5.4,-0.72);
\draw[-,color=white] (5.1,-0.75)                              -- (5.4,-1.025);
\end{tikzpicture}
\caption{Representation of the shape in the unit cell: a 3D representation of the level set sliced by the plane $\phi = 0$ (left) projection of the level set on the Cartesian plane (center), characteristic sets defined by the level set, \textit{i.e.} void and material phases and their reciprocal boundary (right).}
\label{fig:levelset}
\end{figure}

We denote by $Y$ the rescaled unit cell, of coordinates $\vec{y} \in [-1/2,1/2]^2$. 
The strong phase of the micro-architectured unit cell, \textit{i.e.} the shape, is denoted by $\omega$ and is represented by a real-valued auxiliary level set function $\phi$. The principle of the level set method is to implicitly define the interface of a shape via the zero level set of $\phi$ (see equation \eqref{eq:levelset} and \Cref{fig:levelset}). Following the ideas of section 2 in \cite{Allaire2014}, the level set function serves as a base to define the local stiffness tensor $\mathbb{C}(\vec{y})$ in $Y$ as a smooth interpolation between the strong phase and the weak phase, \textit{e.g.} polymer and void material properties, respectively.
\begin{equation}
\displaystyle
\begin{cases}
\begin{aligned}
\phi(\vec{y}) = 0	&	\quad \text{if} \quad \vec{y} \in \partial \omega \cap Y,	&
\quad \text{(boundary) \protect \newboxsymbol{white}{white}} \\
\phi(\vec{y}) > 0	&	\quad \text{if} \quad \vec{y} \in Y \backslash \omega,	&
\quad \text{(void) \protect \newboxsymbol{crazyyellow}{crazyyellow}} \\
\phi(\vec{y}) < 0	&	\quad \text{if} \quad \vec{y} \in \omega,	&
\quad \text{(material) \protect \newboxsymbol{crazyblue}{crazyblue}}
\end{aligned}
\end{cases}
\label{eq:levelset} 		
\end{equation}
\\
The material under consideration occupies a two-dimensional domain $\Omega$, described by a set of coordinates $\vec{x} \in \R^2$ and is modelled as a linear elastic composite with periodic structure. We introduce the small parameter $\epsilon$ as the ratio of the period of the structure to the typical size of the domain $\Omega$ and let $\epsilon \to 0$ to obtain the homogenised problem. The displacement $\vec{u^\epsilon}(\vec{x})$ satisfies the following problem:
\begin{equation}
\begin{aligned}
\vec{\nabla} \cdot \left[ \ten[4]{C}^\epsilon (\vec{x}) \, 
                          \ten[2]{\varepsilon}_{\vec{x}}(\vec{u}^\epsilon) \right] +
                          \vec{f}(\vec{x}) & = \vec{0} \quad \text{in} \, \Omega, \\
\vec{u}^\epsilon (\vec{x}) &= \vec{0} \quad \text{on} \, \partial\Omega.
\end{aligned}
\label{eq:composite}
\end{equation}
where $\mathbb{C}^\epsilon(\vec{x}) \equiv \mathbb{C}(\vec{x} / \epsilon)$ represent a fourth order stiffness tensor which is positive definitive and $\vec{f}(\vec{x})$ is a known body force. Assume that $\vec{u}^\epsilon$ has a two-scale expansion of the form:
\begin{equation}
\vec{u}^\epsilon (\vec{x}) =
\sum_{\alpha=0}^{+\infty} \epsilon^\alpha \, \vec{u}_\alpha \, (\vec{x},\vec{y}), \quad \vec{y} = \frac{\vec{x}}{\epsilon}.
\label{eq:2scale}
\end{equation}
This leads to a series of problems for different orders of $\epsilon$: at order $\epsilon^{-2}$, we obtain that $\vec{u_0}(\vec{x},\vec{y}) \equiv \vec{u_0}(\vec{x})$. At order $\epsilon^{-1}$ we obtain the displacement field solutions of the unit cell problems. At order $\epsilon^0$ we obtain the linear elastic constitutive equation averaged over the unit cell, yielding the following explicit energy formulation of the homogenised elastic tensor $\ten[4]{C}^H$, expressed in terms of its Cartesian components as:
\begin{equation}\label{eq:hom_coef}
C_{ijk\ell}^H = \int_{Y} \ten[4]{C}(\vec{y})
\left( \ten[2]{E}^{i j}   + \ten[2]{\varepsilon}(\vec{\chi}^{i j})   \right) :
\left( \ten[2]{E}^{k\ell} + \ten[2]{\varepsilon}(\vec{\chi}^{k \ell})\right) \, d\vec{y},
\end{equation}
where:
\begin{itemize}[leftmargin=*,topsep=0pt]
\item $\ten[2]{E}^{k\ell}$ designates a constant strain over the unit cell, resulting from
the zero order displacement $\vec{u}_0$. In the 2D case, there are three independent unit strain fields, namely the horizontal unit strain $\ten[2]{E}^{11} = (1,0,0)^T$, the vertical strain $\ten[2]{E}^{22} = (0,1,0)^T$ and the shear unit strain $\ten[2]{E}^{12} = (0,0,1)^T$.
\item $\vec{\chi}^{k\ell}$ represents the displacement fields, solution of the following linear elastic problems with periodic boundary conditions: 
\end{itemize} 
\begin{equation} \label{local:sol}
\begin{aligned}
&\text{Find admissible displacement $\vec{\chi}^{kl}$ such that}\\
&\int_Y \ten[4]{C}(\vec{y})
\left( \ten[2]{E}^{k\ell} + \ten[2]{\varepsilon}(\vec{\chi}^{k\ell}) \right) :
       \ten[2]{\varepsilon}(\vec{v}) \, d\vec{y}= 0,
\end{aligned}
\end{equation}
where $\vec{v}$ are admissible displacement vectors, \textit{i.e.} with zero mean value and adequate smoothness. 

\subsection{Elastic stiffness with orthotropic symmetry}
The effective stiffness tensor $\ten[4]{C}^H$ in \eqref{eq:hom_coef} carries a natural orthotropic material symmetry, provided that $\mathbb{C}$ is isotropic \cite{Sanchez-Palencia1987}. The linear elastic constitutive equation averaged over the unit cell relating the mean stress and strain tensors, denoted as $\ten[2]{\sigma}^H$ and $\ten[2]{\varepsilon}^H$ respectively, has therefore the following expression for the two dimensional problems under consideration:
\begin{equation}
\begin{aligned}
\label{eq:sig:A:eps}
&\ten[2]{\sigma}^H = \ten[4]{C}^H \ten[2]{\varepsilon}^H\\
\text{where:}\qquad &
\ten[2]{\sigma}^H = \mean{\ten[2]{\sigma}}_Y , \qquad
\ten[2]{\varepsilon}^H = \mean{\ten[2]{\varepsilon}}_Y.
\end{aligned}
\end{equation}
In 2D elasticity, the components of $\ten[4]{C}^H$ in matrix notation and in Cartesian coordinates read:
\begin{equation} 
\label{eq:sig:A:eps:coord}
\begin{pmatrix} 
\sigma^H_{11} \\[3pt]
\sigma^H_{22} \\[3pt]
\sigma^H_{12}  
\end{pmatrix}
=
\begin{pmatrix} 
C^H_{1111} & C^H_{1122} & 0 		 \\[3pt]
C^H_{1122} & C^H_{2222} & 0          \\[3pt]
0 		   & 0 			& C^H_{1212} \\[3pt]
\end{pmatrix}
\begin{pmatrix} 
\varepsilon^H_{11} \\[3pt]
\varepsilon^H_{22} \\[3pt]
2 \varepsilon^H_{12}  
\end{pmatrix}
\end{equation}
Alternatively, one could express the effective strain as a function of the effective stress with the following effective material tensor:
\begin{equation}
\begin{pmatrix} 
\varepsilon^H_{11} \\[3pt]
\varepsilon^H_{22} \\[3pt]
2 \varepsilon^H_{12}  
\end{pmatrix}
=
\begin{pmatrix} 
1/E_1 & -\nu_{12}/E_2 & 0 \\[3pt]
-\nu_{21}/E_1 & 1/E_2 & 0 \\[3pt]
0 & 0 & 1/G \\[3pt]
\end{pmatrix}
\begin{pmatrix} 
\sigma^H_{11} \\[3pt]
\sigma^H_{22} \\[3pt]
\sigma^H_{12}  
\end{pmatrix}
\end{equation}
where $E_i$ denote the homogenised Young moduli, $\nu_{ij}$ denote the Poisson's ratios and $G$ denotes the homogenised shear modulus. Let us further remark, that by symmetry of the elastic compliance matrix, the following ratios have to be equal:
\begin{equation}
\frac{\nu_{12}}{E_2} = \frac{\nu_{21}}{E_1}
\end{equation}
The elastic moduli, $C^H_{ijkl}$, can equally be expressed in terms of the compliance moduli, \textit{i.e.} Young moduli and Poisson's ratios: $C^H_{1111}= (1-{\nu_{12}\nu_{21}})^{-1}E_1$, $C^H_{2222}= (1-{\nu_{12}\nu_{21}})^{-1}E_2$, $C^H_{1122}=\nu_{21}(1-{\nu_{12}\nu_{21}})^{-1}E_1$, $C^H_{2211}=\nu_{12}(1-{\nu_{12}\nu_{21}})^{-1}E_2$ with $C^H_{1122}=C^H_{2211}$ as can be easily obtained from the inversion of the corresponding matrices. A simple calculation immediately yields:
\begin{equation}
\nu_{12}=\frac{C^H_{1122}}{C^H_{2222}} \text{ and } \nu_{21}=\frac{C^H_{1122}}{C^H_{1111}}.
\label{eq:APR_stiffness_comp}
\end{equation}
Moreover, the homogenised Poisson's ratio $\nu_{ij}$ are equally denoted \textit{effective} Poisson's ratio to highlight their reference to the homogenised unit cell. For example $\nu_{12}$ characterises the contraction of the structure in the direction of $Oy$ axis when the cell stretched in the direction of $Ox$ axis and in general $\nu_{12} \ne \nu_{21}$. However, if the micro-architecture of the unit cell obeys ``cubic'' symmetry we have $C^H_{1111} = C^H_{2222}$ and we trivially obtain that $E_1=E_2=E^*$ and $\nu_{12} = \nu_{21}=\nu^*$. 

\subsection{Shape optimisation of the micro-structures}
Next we discuss the framework of the optimisation problem without presenting the details of the algorithm which is beyond the scope of the paper. A detailed discussion is given by the authors in \cite{Nika2019}.\smallskip\\
Assume that the unit cell $Y$ is the working domain and consider $\omega$ an open and bounded subset of $Y$ representing the shape of the micro-architecture. The optimisation problem seeks to find the shape $\omega$ such that the effective stiffness of the material reaches a given target value $\mathbb{C}^t$. The problem can be formulated as constrained minimisation problem. The distance between the effective elastic moduli over the unit cell and target elastic moduli can be measured by the following cost functional: 
\begin{equation}
\label{eq:objective}
\mathcal{J}(\omega) = \frac{1}{2} \norm{\mathbb{C}^H(\omega) - \mathbb{C}^t}^2_{\eta}
\end{equation}
where $\norm{\cdot}_{\eta}$ is the weighted Euclidean norm, $\mathbb{C}^t$ is the target elastic tensor, and $\eta$ are the weight coefficients. We define a set of admissible shapes contained in the working domain $Y$ and have a prescribed volume by $\mathcal{U}_{ad} = \left \{ \omega \subset Y \text{ such that } |\omega| = V^t \right \}.$ Hence, the optimal shape design of the micro-architecture can be formulated as the following optimisation problem: 
\begin{gather}
\begin{aligned} \label{opti:prob}
& \inf_{\omega \subset \mathcal{U}_{ad}}  \mathcal{J}(\omega), \\
\vec{\chi}^{kl} & \text{ satisfies } \eqref{local:sol}.
\end{aligned}
\end{gather} 
In practice, the volume constraint is enforced using a Lagrange multiplier (the technique for updating the Lagrange multiplier is based upon the works of Allaire \& Pantz \cite{Allaire2006}).

\subsection{Numerical algorithm}
\begin{figure}[h]
\centering
\begin{tabular}{cccc}
\subf{\includegraphics[width=32mm]{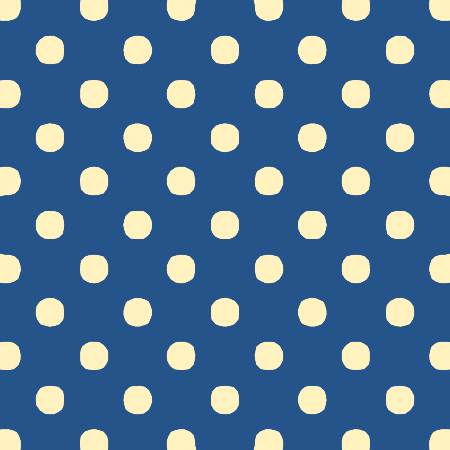}}{(a)} &
\subf{\includegraphics[width=32mm]{ex3_micro.pdf}}{(b)} &
\subf{\includegraphics[width=32mm]{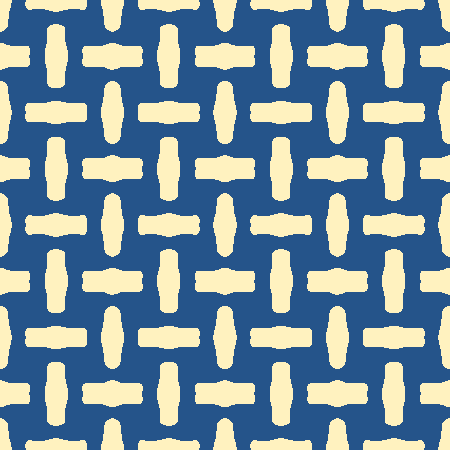}}{(c)} &
\vbox{\hbox{\strut \newboxsymbol{crazyblue}{crazyblue} Material}\hbox{\strut \newboxsymbol{crazyyellow}{crazyyellow} Void}}
\end{tabular}
\caption{The design process of the material. Image (a) depicts the initial guess of the micro-structure, (b) depicts the optimal micro-structure after $100$ iterations, (c) depicts the macroscopic material assembled periodically from the optimal unit cell of image (b).}
\label{example3}
\end{figure}

The optimisation of $\mathcal{J}(\omega)$ is carried out by advecting an initial shape $\omega_0$ with velocity $v$ obtained from the shape derivative $\dpair{\mathcal{J}'(\omega)}{\vec{\theta}}$ in the direction $\vec{\theta}$ (see Allaire, Jouve \& Toader \cite{Allaire2004} or Wang, Mei \& Wang \cite{Wang2004}). The advection is realised by solving the Hamilton-Jacobi equation:
\begin{equation}
\phi_{,t} + v \, |\vec{\nabla} \phi| = 0
\end{equation}
where $v$ is the velocity of the interface computed from the shape derivative $\dpair{\mathcal{J}'(\omega)}{\vec{\theta}}$.\smallskip\\
The numerical algorithm can be summarised in the following steps:
\begin{itemize}[leftmargin = 24pt]
\item[(i)] Initialise the level set $\phi_0$ corresponding to the initial shape $\omega_0$.
\item[(ii)] Update the level set $\phi_0$ using the signed distance function $d_{\omega_0}$.
\item[(iii)] Iterate until convergence for $k \ge 0$:
\begin{itemize}
	\item[a.] Calculate the local solutions $\vec{\chi}^{m\ell}_k$ for $m,\ell=1,2$ by solving the linear elasticity problem in $Y$. 
	\item[b.] Deform the domain $\omega_k$ by solving the above Hamilton-Jacobi equation. The new shape $\omega_{k+1}$ is characterised by the level set $\phi_{k+1}$ after a time step $\Delta t_k$. The time step $\Delta t_k$ is chosen so that $\mathcal{J}(\omega_{k+1}) \le \mathcal{J}(\omega_k)$. 
\end{itemize}
\item[(iv)] If needed for stability reasons, re-initialise the level set functions $\phi_k$.
\end{itemize}
The complete algorithm as well as several examples are presented in Nika \& Constantinescu \cite{Nika2019}. Additional mathematical results and algorithmic issues can be found in the works of Allaire, Jouve \& Toader \cite{Allaire2004}, Allaire et al. \cite{Allaire2014}, Wang, Mei \& Wang \cite{Wang2004} for more details about the mathematical results and algorithmic issues for the solution method. Let us remark that the algorithm does not allow for nucleations of voids. However the level set method is well known to handle easily topology changes, \textit{i.e.} merging or cancellation of holes. Therefore, algorithm is able to perform topology optimisation if the number of holes of the initial design is sufficiently large and converges smoothly to a (local) minimum which strongly depends on the initial topology. Next, for each numerical example, we provide the initials guesses for the shapes, which are typically a plate filled with holes, as illustrated in \Cref{example1}(a), \ref{example2}(a) and \ref{example3}(a). 

\subsection{Examples of obtained microstructures} \label{sec:example}
\begin{table}[b]
\centering
\begin{tabularx}{\textwidth}{c|*{5}{>{\centering \arraybackslash}X}}
Example & $\mathbb{C}^t$ & $\mathbb{C}^H(\omega)$ & $\nu^t$ & $\nu^*$ & Shape $\omega$\\
\hline
1 &
{$\displaystyle
	\begin{pmatrix}
    \phantom{-}0.1 & -0.1 & 0 \\
    -0.1 & \phantom{-}0.1 & 0 \\
   	0 & 0 & G \\
  	\end{pmatrix}$} &
{$\displaystyle
	\begin{pmatrix}
    \phantom{-}0.12 & -0.05 & 0 \\
    -0.05 & \phantom{-}0.04 & 0 \\
   	0 & 0 & 0.006 \\
  	\end{pmatrix}$} &
$-1.$ &
$\{-1.25, -0.42\}$ &
\vspace{0.1cm}
\includegraphics[width=0.1\columnwidth]{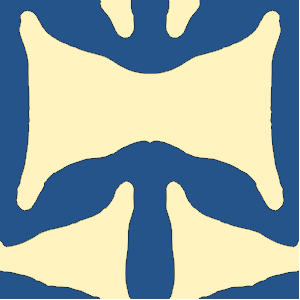}\\
2 &
{$\displaystyle
	\begin{pmatrix}
    \phantom{-}0.1 & -0.1 & 0 \\
    -0.1 & \phantom{-}0.1 & 0 \\
   	0 & 0 & G \\
  	\end{pmatrix}$} &
{$\displaystyle
	\begin{pmatrix}
    \phantom{-}0.12 & -0.05 & 0 \\
    -0.05 & \phantom{-}0.12 & 0 \\
   	0 & 0 & 0.003 \\
  	\end{pmatrix}$} &
$-1.$&
$-0.42$&
\vspace{0.1cm}
\includegraphics[width=0.1\columnwidth]{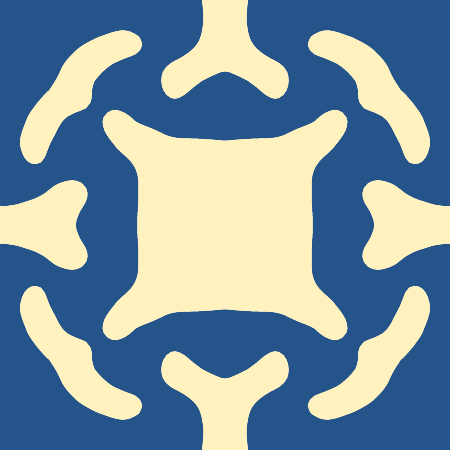}\\
3 &
{$\displaystyle
	\begin{pmatrix}
    \phantom{-}0.2 & -0.1 & 0 \\
    -0.1 & \phantom{-}0.2 & 0 \\
   	0 & 0 & G \\
  	\end{pmatrix}$} &
{$\displaystyle
	\begin{pmatrix}
    \phantom{-}0.19 & -0.09 & 0 \\
    -0.09 & \phantom{-}0.19 & 0 \\
   	0 & 0 & 0.06 \\
  	\end{pmatrix}$} &
$-0.5$&
$-0.47$&
\vspace{0.1cm}
\includegraphics[width=0.1\columnwidth]{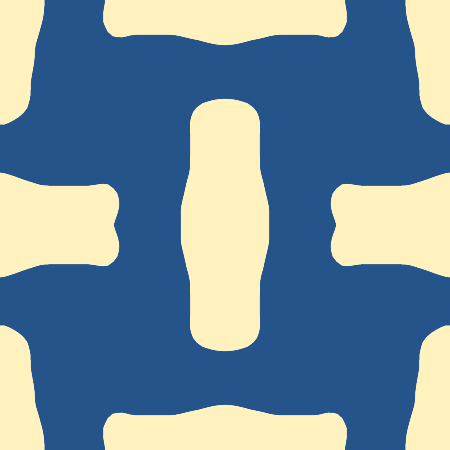}\\
\end{tabularx}
\caption{From left column to right column: target elastic stiffness tensor $\mathbb{C}^t$, final elastic tensor $\mathbb{C}^H(\omega)$, target Poisson's ratio $\nu^t$, final effective Poisson's ratio $\nu^*$ unit cell shape $\omega$ for the discussed examples. Let us remark that the bottom two structures on the right column carry cubic symmetry, while the top structure on the right column carries orthotropic symmetry.}
\label{table:Hcoef}
\end{table}
In all the examples that follow, the Young's modulus was set to to $E_m = 0.91$ MPa for the strong phase (material) and $E_v = 0.91 \times 10^{-3}$ MPa for the weak phase (void). The Poisson's ratio was set to $\nu = 0.3$ for both phases. Under the plane stress assumption, the components of the elastic tensor of the base material are $C_{1111}^{m} = C_{2222}^{m} = 1.0 \text{ MPa ; } C_{1122}^{m} = 0.3 \text{ MPa ; } C_{1212}^{m} = 0.35\text{ MPa}$. The  unit cell $Y$ was meshed with a structured symmetric grid of $100 \times 100$ quadrangular each formed of four equal triangular linear elements ($P1$). All computations were carried out using an in house programming of the preceding algorithm \cite{Nika2019} operating on {\tt FreeFEM++} software \cite{Hecht2012}. The optimisation is assumed to be terminated when 100 iterative steps are reached.\smallskip\\
The main intention of the present work was to design micro-architectured materials exhibiting an effective negative Poisson's ratio. However, in all examples the target objective was defined only in terms of the coefficients $C^H_{1111}, C^H_{1122}, C^H_{2222}$ using relation \eqref{eq:APR_stiffness_comp}. The shear coefficient $C^H_{1212}$ as well as the $C^H_{1211}$ and $C^H_{1222}$ coefficients were left free. Therefore, only the elastic moduli of the unit cell corresponding to the direction $11$ and $22$ directions of strain and stresses were controlled. 

\begin{figure}[h]
\centering
\begin{tabular}{cccc}
\subf{\includegraphics[width=32mm]{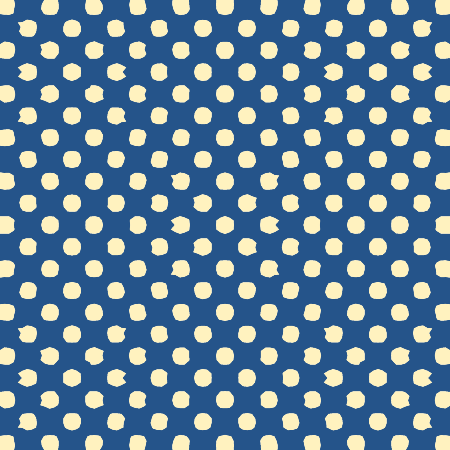}}{(a)} &
\subf{\includegraphics[width=32mm]{ex1_micro}}{(b)} &
\subf{\includegraphics[width=32mm]{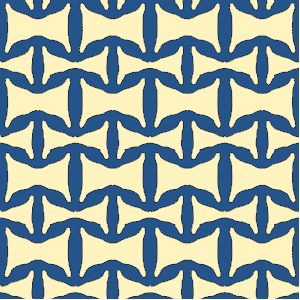}}{(c)} &
\vbox{\hbox{\strut \newboxsymbol{crazyblue}{crazyblue} Material}\hbox{\strut \newboxsymbol{crazyyellow}{crazyyellow} Void}}
\end{tabular}
\caption{Initial and final shape of the microstructures: (a) initial shape consisting of a series of circular micro-perforations (b) final, optimal, shape of the unit cell after $100$ iterations (c) final, optimal, shape of the periodic material.}
\label{example1}
\end{figure} 
\paragraph{Example 1.} The first micro-structure to be optimised is a structure whose target effective Poisson's ratio is equal $\nu^t = -1.0$. 
The volume constraint was set to $V^t=50\%$. We further note that for this structure we enforced a symmetry of the shape along the vertical axis, by symmetrising the level set function after each iteration in the algorithm. The initial and final shape of the micro-structure on the unit cell and as a periodic material are represented in \Cref{example1}.\smallskip\\
The final shape can be characterised as an re-entrant honeycomb structure and looks similar to the designs imagined by Almgren \cite{Almgren1985}. Its homogenised coefficients, displayed in \Cref{table:Hcoef}, show that the structure exhibits an effective orthotropic behaviour and a simple calculation yields $\nu_{12} = -1.25$ and $\nu_{21}=-0.42$. Hence, the expansion of the structure along the $Oy$ axis when stretched in the $Ox$ axis is larger than the expansion along the $Ox$ axis when stretched in the $Oy$ axis. This non-symmetric effect has been enabled as the symmetry relation was only imposed along the $Oy$ axis in the algorithm.\smallskip
\begin{figure}
\centering
\includegraphics[width=0.5\textwidth]{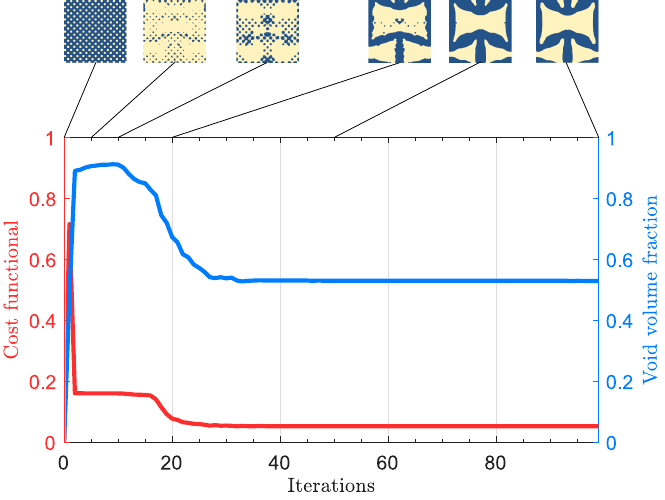}
\caption{Evolution of the cost functional (in red) and volume constraint V (in blue) during $100$ iterations.}
\label{example1:convergence}
\end{figure}%
\\
The convergence history of the cost-functional and of the volume constraint displayed in \Cref{example1:convergence} shows that the target coefficient got stabilised in slightly more than 20 iteration and that the later iteration contributed only to small improvements without bringing the cost functional to less than $0.06$ which corresponds to $92\%$ decrease of the initial value. The gap with respect to the target moduli can be read from \Cref{table:Hcoef}. It is interesting to remark, that the final optimised micro-structure has a shear moduli close to $0$. However, the final effective Poisson's ratio is close to the set target as will be discussed in the comparison with the printed samples. The volume constraint has a different evolution than the cost-functional with an initial increase given by the initial evolution of the holes and then a fast and a slow evolution which lies within the proposed range of the constraint.

\begin{figure}[h]
\centering
\begin{tabular}{cccc}
\subf{\includegraphics[width=32mm]{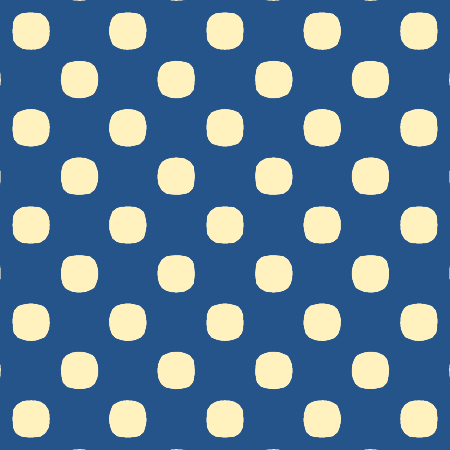}}{(a)} &
\subf{\includegraphics[width=32mm]{ex2_micro.pdf}}{(b)} &
\subf{\includegraphics[width=32mm]{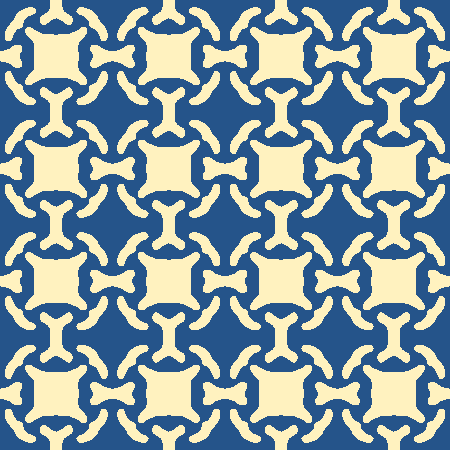}}{(c)} &
\vbox{\hbox{\strut \newboxsymbol{crazyblue}{crazyblue} Material}\hbox{\strut \newboxsymbol{crazyyellow}{crazyyellow} Void}}
\end{tabular}
\caption{The design process of the material. Image (a) depicts the initial guess of the micro-structure, (b) depicts the optimal micro-structure after $100$ iterations, (c) depicts the macroscopic material assembled periodically from the optimal unit cell of image (b).}
\label{example2}
\end{figure}
\paragraph{Example 2.} For the second micro-structure to be optimised, the target effective Poisson's ratio was also $\nu^t=-1$. The target tensor possesses a cubic symmetry, meaning the desired mechanical properties along the $Ox$ and $Oy$ axis should be equal. This time, the void volume fraction constraint was of an inequality type, and was set to $16\% \leq V^t \leq 60\%$. To counter the loss of symmetry observed in the previous example, a symmetry of the shape was enforced along both the $Ox$ axis and $Oy$ axis, by symmetrising the level set function during the algorithmic iterations. The initial and final shape of the micro-structure on the unit cell and as a periodic material are represented in \Cref{example2}, while the target and final elastic moduli are shown in \Cref{table:Hcoef}.\smallskip\\
As prescribed, the resulting structure exhibits a ``cubic'' symmetry. The computed effective Poisson's ratio is $\nu^*=\nu_{12}=\nu_{21}=-0.42$. By comparing at the target and the obtained elastic tensor, one can remark that the diagonal elastic moduli $C^H_{1111}$ and $C^H_{2222}$ are fairly close to the target but the shear $C^H_{1122}$ fails at attaining the desired properties of $\nu^t=-1$. This suggests that there is a trade-off between symmetrical tensor and extreme negative Poisson's ratio in the optimisation algorithm. This will further be discussed in the following section. Finally, let us remark that the final volume ratio is at approximately $36\%$ and lies in the middle of the imposed interval. As before, the final optimised micro-structure has a shear modulus $C^H_{1212}$ close to $0$.

\paragraph{Example 3.} For the third micro-structure to be optimised, the target effective Poisson's ratio was $\nu^t=-0.5$. The target tensor possesses a cubic symmetry, meaning the desired mechanical properties along the $Ox$ and $Oy$ axis should be equal. The void volume fraction constraint is set to $V^t = 40\%$ and is updated the same way as in the first example. As in example 2, a symmetry of the shape was enforced along the $Ox$ and $Oy$ axis.\smallskip\\
The final shape is of the rotating units type, discussed in Grima et al. \cite{Grima2007}. As prescribed, the resulting structure exhibits a ``cubic'' symmetry. The computed effective Poisson's ratio is $\nu^*=\nu_{12}=\nu_{21}=-0.47$. By comparing the target and the obtained elastic tensor, one can remark that the final elastic moduli are fairly close to the target and that this structure has a shear modulus $C^H_{1212}$ which is of the same order of magnitude as the other moduli. Finally, let us remark that the final volume ratio is at approximately $43\%$ and lies in the vicinity of the imposed value.

\subsection{Representation of the examples in the space of elastic stiffness}
\begin{figure}[b]
\centering
\includegraphics[scale=0.75]{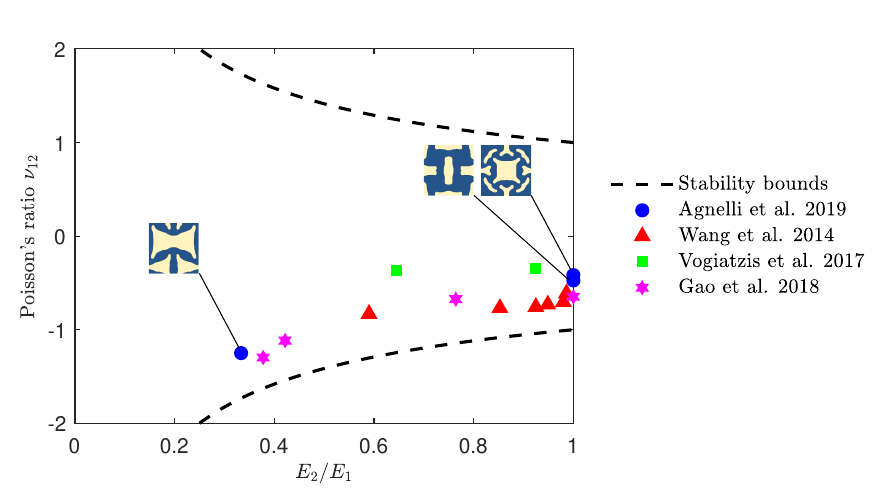}
\caption{Optimal shapes obtained from this article as well as from the literature \protect \cite{Wang2014,Vogiatzis2017,Gao2018} plotted against the stability bounds. Remark that extreme Poisson's ratio (\textit{i.e.} below $-1$) are reachable for anisotropic structures.}
\label{fig:Stability}
\end{figure}

The elastic tensor, $\mathbb{C}^H$, governs the overall effective material response to an applied load. Since we assumed that the material tensor characterising the microstructure is positive definite then $\mathbb{C}^H$ is also positive definite. In terms of the physical parameters $E_1,E_2, \nu_{12}, \nu_{21}$ the positive definiteness of the effective material tensor requires that the stiffness tensor is positioned with the following {\it stability} bounds (see the works of Ting \& Chen \cite{Ting2005} for a detailed discussion):
\begin{equation}
|\nu_{12}| \le \sqrt{\frac{E_1}{E_2}}, \quad |\nu_{21}| \le \sqrt{\frac{E_2}{E_1}}.
\label{Thm:bounds}
\end{equation}
In \Cref{fig:Stability} we plot the Poisson's ratios against the {\it stability} bounds of certain optimal shapes from some recent articles published and see how they compare with our own optimised microstructures. A material that would have $E_1 = E_2$ and $\nu_{21}=\nu_{12}=\nu^*=-1$ would fall on the lower \textit{stability} bound. From the graph of \Cref{fig:Stability}, we understand that the structure has to loose its ``cubic'' symmetry and accept an important stiffness unbalance between the directions 1 and 2, expressed by the ratio $E_2 / E_1$ to reach one extreme negative Poisson's ratio. \smallskip\\
The shapes from the literature, that have been designed by the optimisation of a level set function, share generic features: lattice architectures with re-entrant corners (like re-entrant honeycomb) or rotating semi-rigid units connected by flexible hinges. Moreover, the orthotropic structures with $E_2 / E_1 < 0.8$ are only lattice architectures with re-entrant corners. Let us further mention the recent contributions of \cite{Zhang2018} who designed a chiral bi-material micro-architecture with $E_2 / E_1 \approx 0.65$ and an effective Poisson's ratio $\nu_{12}=-0.8$ using the SIMP optimisation method, which would also fill in the data cloud of \Cref{fig:Stability}. Lastly, the examples from literature, aimed to reach a maximal negative Poisson's ratio, fall at a certain distance from the inferior stability bound. We therefore understand the need for tighter bounds to be used as a guide to effectively explore the ability of the algorithm to ascertain which elastic moduli are attainable in future designs.


\section{Analysis of fabricated polymer structures}
\subsection{Manufacturing process: equipment and materials} 
\begin{figure}[h]
\centering
\includegraphics[scale=0.75]{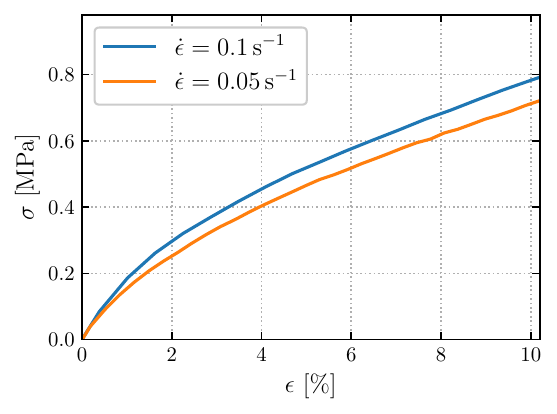} 
\caption{Base material response to uniaxial tensile loading. Homogeneous stress-strain curves.}
\label{GM08B}
\end{figure}

\begin{figure}
\centering
\begin{tabular}{ccc}
\subf{\includegraphics[width=.413\columnwidth]{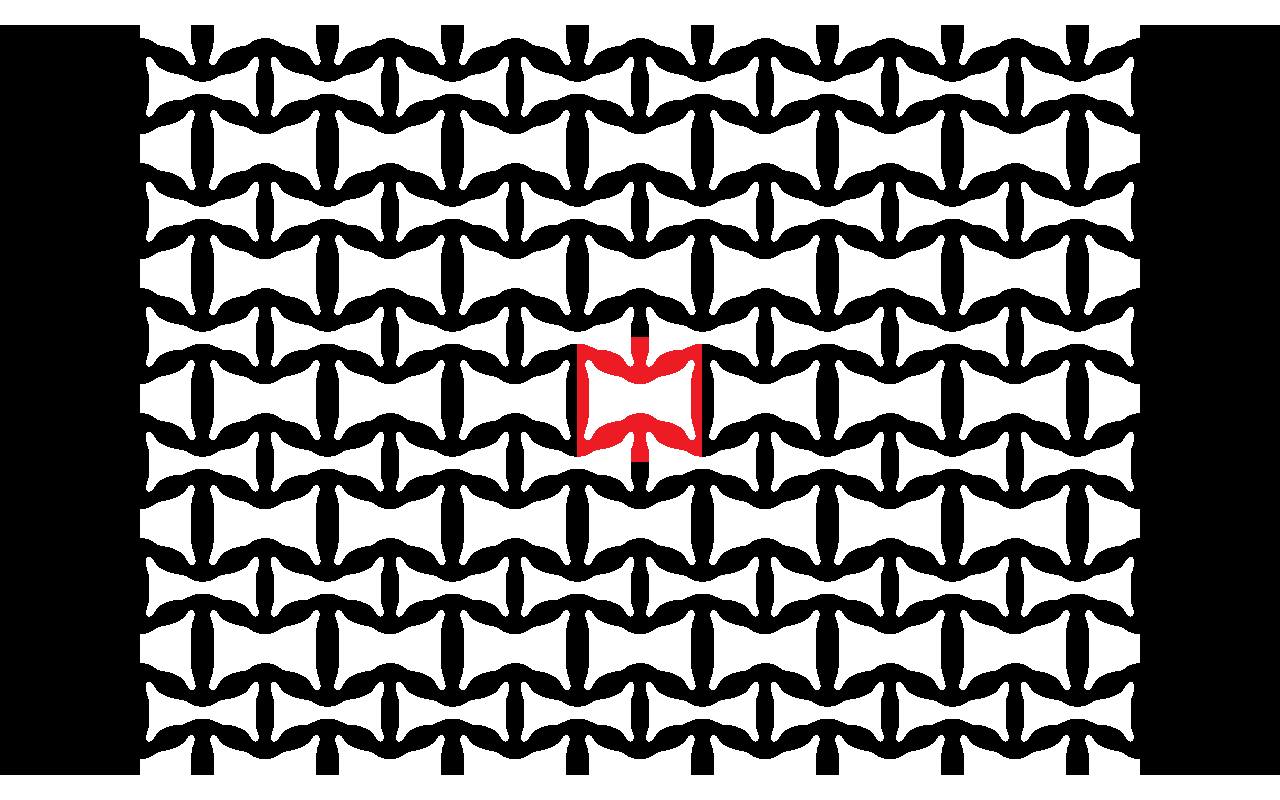}}
     {Example 1: Input design}
&
\subf{\includegraphics[width=.45\columnwidth]{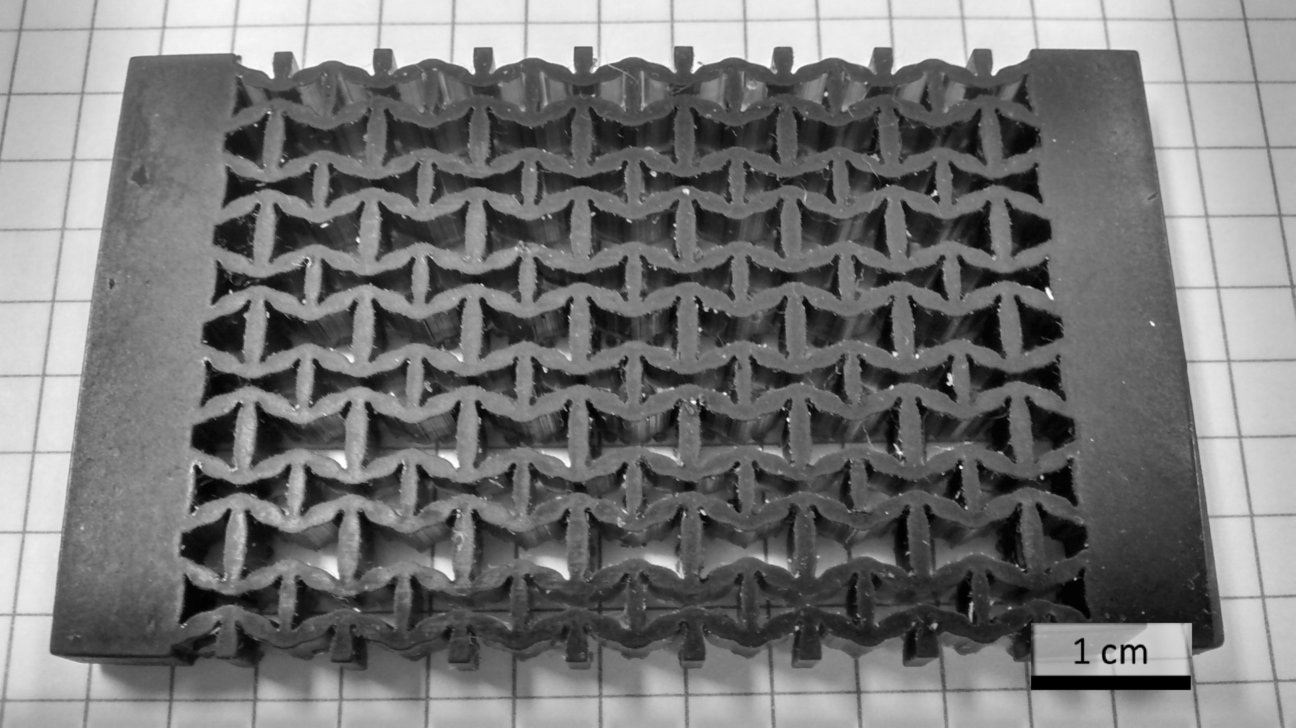}}
     {Example 1: Fabricated result}
     \\
\subf{\includegraphics[width=.405\columnwidth]{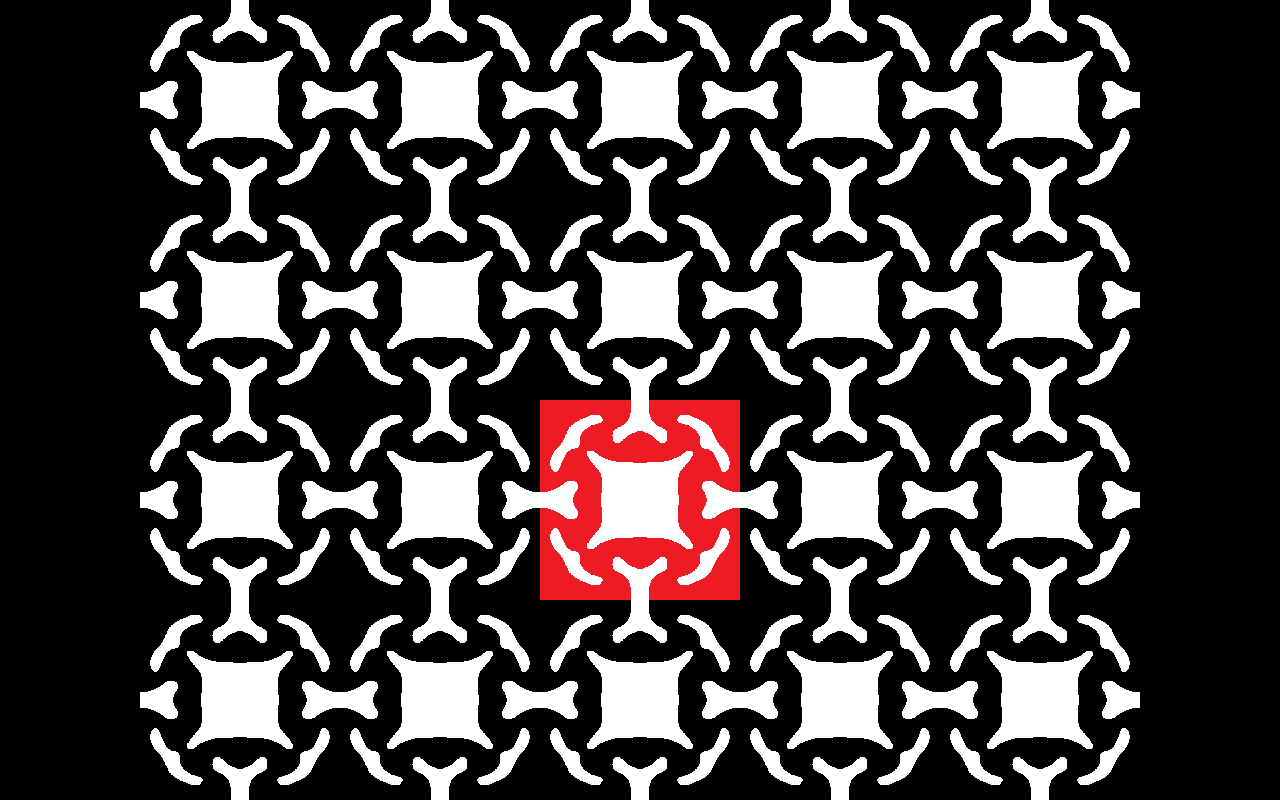}}
     {Example 2: Input design}
&
\subf{\includegraphics[width=.45\columnwidth]{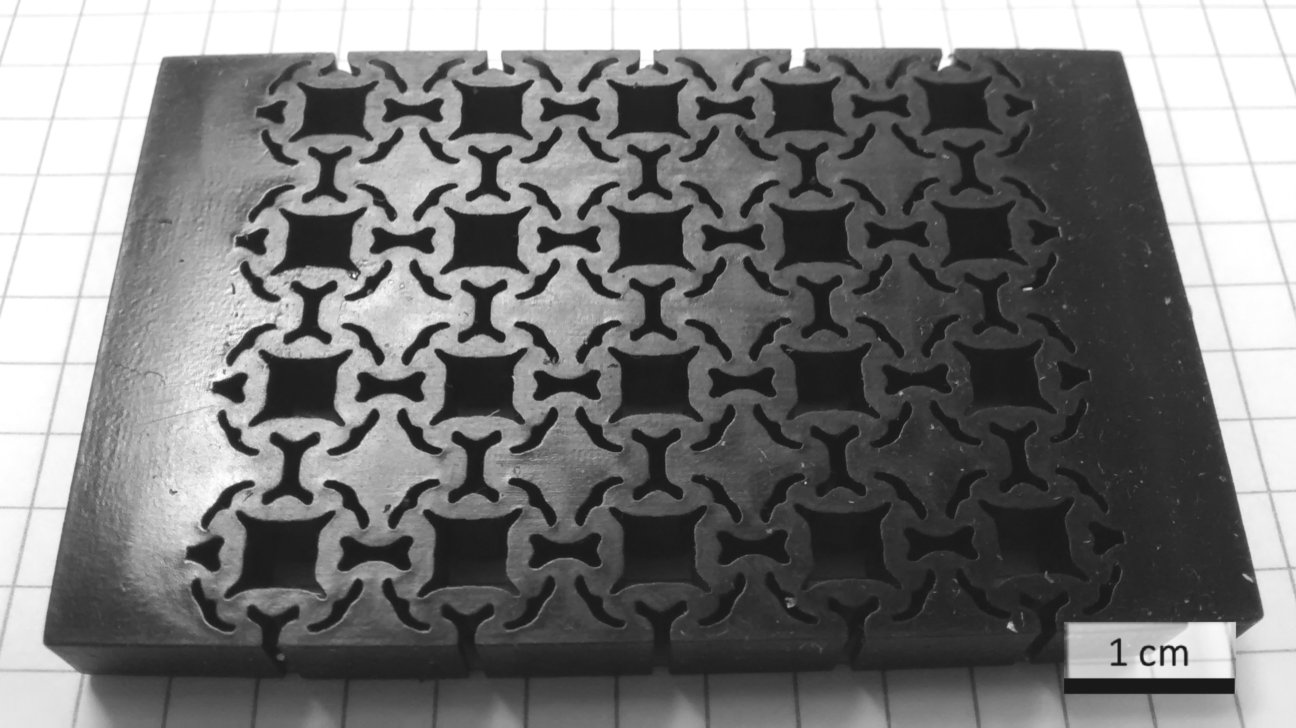}}
     {Example 2: Fabricated result}
     \\
\subf{\includegraphics[width=.41\columnwidth]{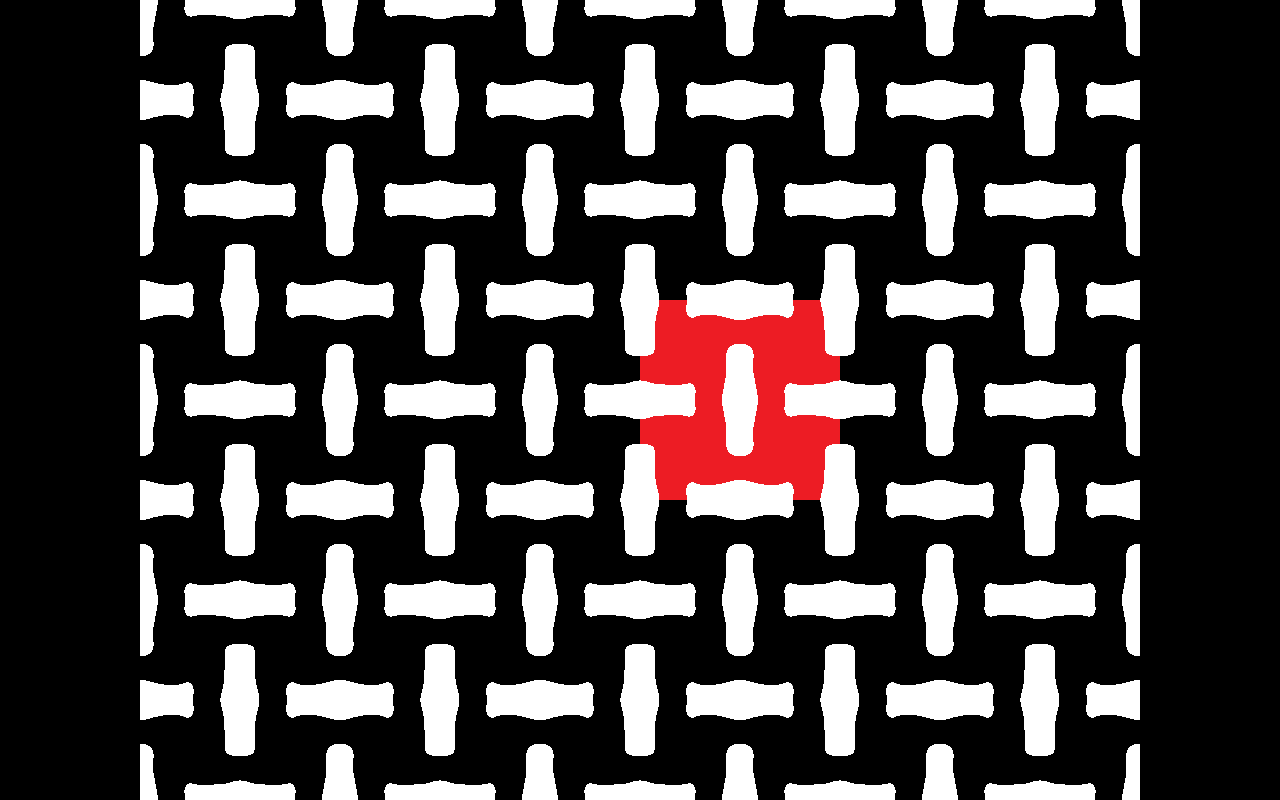}}
     {Example 3: Input design}
&
\subf{\includegraphics[width=.45\columnwidth]{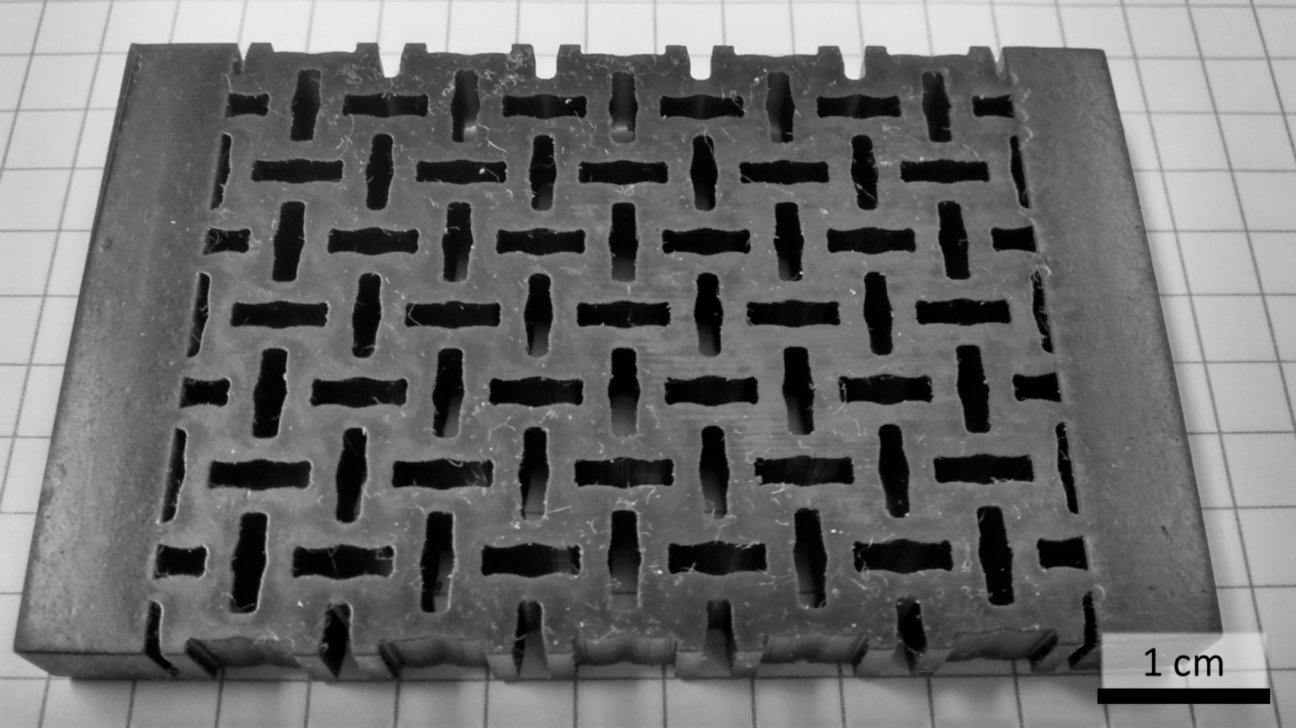}}
     {Example 3: Fabricated result}
\end{tabular}
\caption{Fabricated specimen from the examples of section 2. Digital image fed into the 3D printer (left) and final printed specimen (right). The red coloured unit cells were the cells observed during the digital image correlation measurements.}
\label{fig:3D}
\end{figure}

The optimal shapes have been additively manufactured with digital light processing stereo-lithography technology (DLP) using a commercial Ember 3D printer. A digital projector screen flashes a single image of each layer across the entire surface of the vat filled of photo-sensitive liquid resin at once, causing chains of molecules to link and thus forming solid polymer. The process is repeated until the 3D model is complete. Then the vat is drained of liquid, revealing the solidified model and the solid model is washed with a solvent.\smallskip\\
The printer has a resolution of $50 \, \mathrm{\mu m}$, corresponding to 1 pixel in the digital projector screen, and a range of the processing layer thickness of $10-100 \, \mathrm{\mu m}$. The largest processing build volume is $64 \, \mathrm{mm} \times 40 \, \mathrm{mm} \times 134 \, \mathrm{mm}$ (note that $64 \, \mathrm{mm} \times 40 \, \mathrm{mm}$ correspond to a $1280 \times 800$ pixels picture). For a thickness of $25 \, \mathrm{\mu m}$ per layer, the speed range is of $18 \, \mathrm{mm.h^{-1}}$. The printable minimal feature size of the specimens is announced at $0.4 \, \mathrm{mm}$ corresponding roughly to $8$ pixels.\smallskip\\
We selected a rubber-like material, commercially denoted as GM08b\footnote{Characteristics of this material can be found in the manufacturers data sheet (see \texttt{https://dl.airtable.com})}, as the base material because of its compliant nature. \Cref{GM08B} displays a representative tensile stress--strain curve of this material. As expected for a rubber-like material it does not display an ideal linear elastic behaviour, it exhibits a gradually variation of the stiffness with increasing strain.
\\
The optimal shapes obtained in examples 1--3 (see \Cref{example1}, \ref{example2} and \ref{example3}) are represented by the final level set function. The later presents a smooth variation between values corresponding to the two materials in a neighbourhood of their interface and therefore the level set representation has been binarised and extruded in the $Oz$ direction in order to create a three-dimensional object. More precisely, the 3D printed samples have been produced by the following procedure:
\begin{itemize}[leftmargin = 24pt]
\item[(i)] Binarize the level set function obtained by shape optimisation.
\item[(ii)] Create a periodic array for each sample: $8 \times 6$ unit cells for Example 1, $5 \times 4$ unit cells for Example 2 and Example 3. The final result was a binarized $1280 \times 800$ pixels image (see \Cref{fig:3D} for details). 
\item[(iii)] Extrude the preceding image to obtain the 3D sheet of the desired height. The dimensions of the printed samples are $64 \, \times 38 \, \times 6 \, \mathrm{mm}$ for Example 1 and $64 \times 40 \, \times 6 \, \mathrm{mm}$ for Example 2 and 3.
\item[(iv)] Print the files with the following processing parameters: laser power was $5 \,\mathrm{W}$, the exposure time $1\,\mathrm{s}$ per layer and the layer thickness was $50 \, \mathrm{\mu m}$.
\item[(v)] Wash the samples in an isopropanol bath for $5\,\mathrm{min}$.
\item[(vi)] Post-cure the samples for $30\,\mathrm{min}$ in an UV oven at $2000\,\mathrm{W}$.
\end{itemize}

\subsection{Testing and full-field displacement measurement using Digital Image Correlation (DIC)} 
A series of uniaxial static tensile tests were undertaken to assess the tensile properties of the auxetic lattice structures by using a home-made testing machine with a symmetric displacement of the two cross-heads and equipped with a $100 \, \mathrm{N}$ load cell. The tensile tests were performed at a rate of $0.05 \, \mathrm{mm.s^{-1}}$ up to $3\, \mathrm{mm}$, which corresponds to a strain rate of $\dot{\varepsilon} = 10^{-3} \, \mathrm{s^{-1}}$ up to a maximal strain of $\varepsilon = 5\%$.\smallskip\\
The tensile tests were recorded and used for full-field measurements by digital image correlation (DIC). The recordings were obtained using a high-resolution digital camera (Schneider Optics 8-bit camera with a Makro-UNIFOC 100/77 lens) mounted on the tensile testing machine and grey scale pictures resolution of $4904 \times 3280$ were recorded every second during the loading. The camera was mounted on a perpendicular axes with respect to the plane of the specimen, which enables the direct use of a 2D DIC. To improve the precision of the measurement, a white speckle pattern was placed on the sample by airbrushing.\smallskip\\
The DIC was performed using the CorrelManuV 2D (CMV) software, developed by M. Bornert  \cite{Allais1994}. The processed displacement field corresponds to a single unit cell in the middle of the structure at five different loading time steps, using a $100 \times 100$ grid, \textit{i.e.} having $10 000$ measurement points. For each node, the subset size was set to $20 \times 20$ pixels, while the searching area was set to $100 \times 100$ pixels. The measurement included a computation without transformation, \textit{i.e.} rotation of the subset window and a re-optimisation allowing transformations with a reduced searching area of $30 \times 30$ pixels.

\subsection{Experimental results}
\begin{figure}
\centering
\subf{\includegraphics[width=0.45\textwidth]{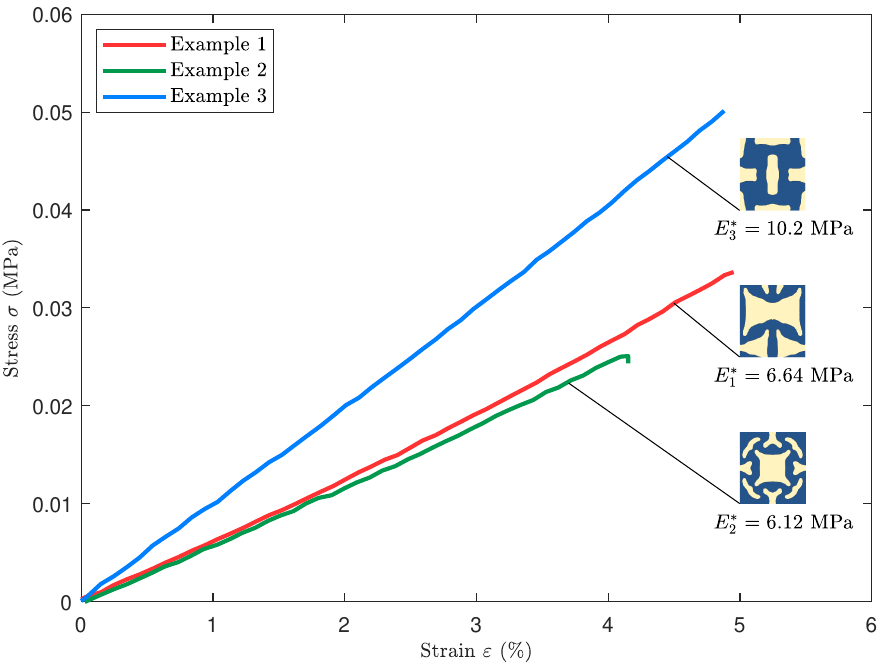}}{(a)}
\hspace{1cm}
\subf{\includegraphics[width=0.45\textwidth]{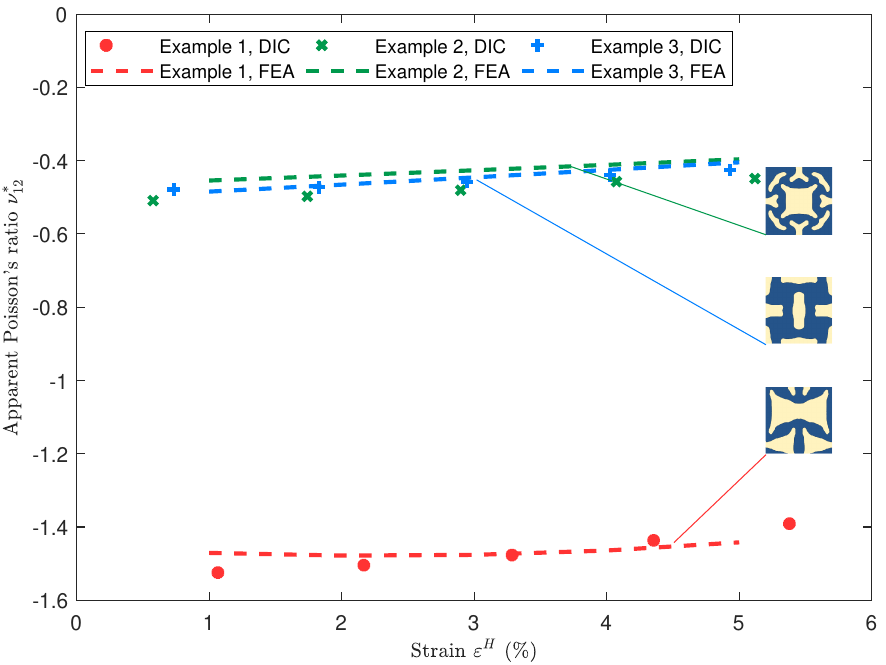}}{(b)}
\caption{(a) Effective stress--strain curves for all three examples overlaid in the same plot obtained experimentally by performing a uniaxial tensile test. One can clearly observe all three structures exhibit linear behaviour for strains up to $5\%$. (b) Comparison of the evolution of the Poisson's ratio plotted as a function of the effective strain from measurements taken by digital image correlation versus computations made by finite element analysis. We can clearly observe a trend on all three materials of the loss of their auxeticity as the uniaxial strain increases beyond the $5\%$ mark.}
\label{APR}
\end{figure}
\begin{figure}
\centering
\includegraphics[width=0.8\textwidth]{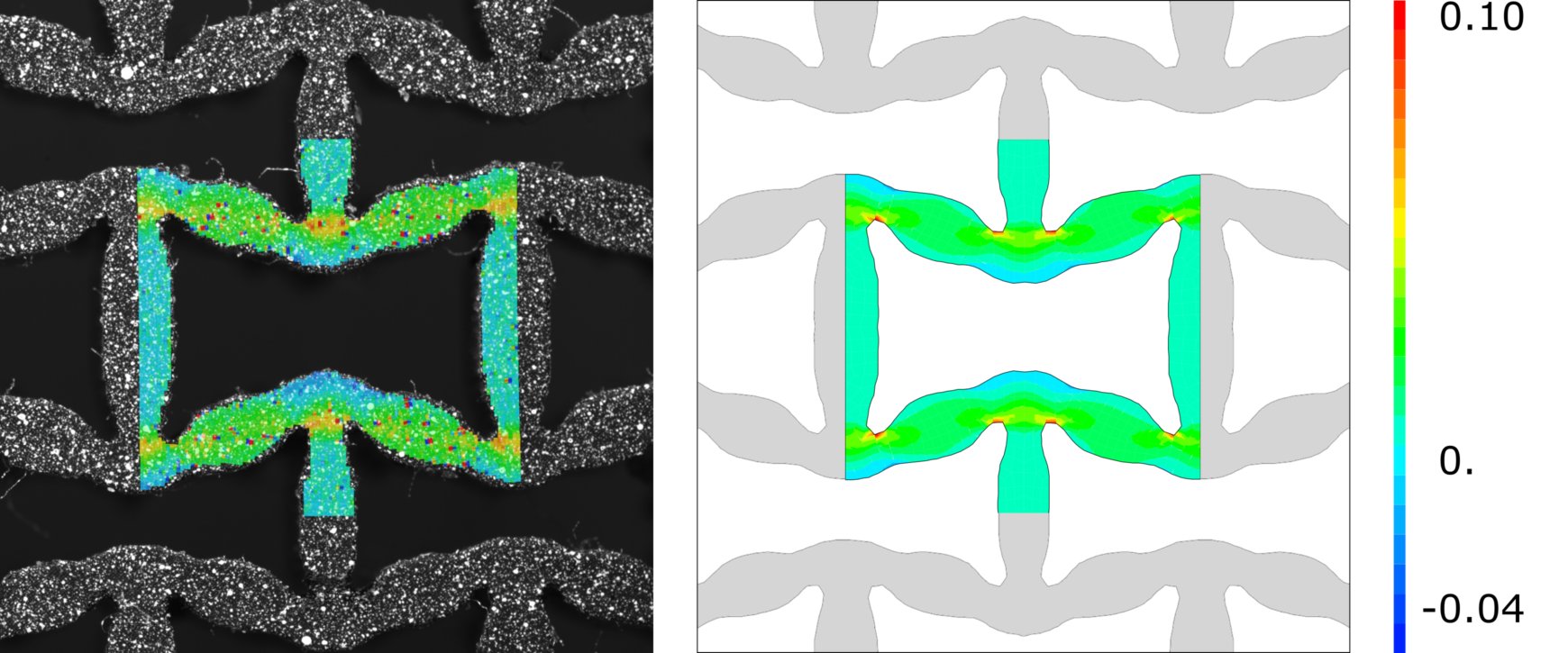}
\caption{Plot of the dimensionless values of the horizontal tensile strain field in a single unit cell of the optimised structure. The strain field for the image on the left was computed from the raw displacement field obtained by Digital Image Correlation, while for the image on the right using finite element analysis. We remark that our DIC approach generated some outliers that shall not be taken into account.}
\label{fig:strain}
\end{figure}

The stress strain response under a uniaxial tensile test along $Ox$ for the three materials are displayed in \Cref{APR}(a). One can observe a linear behaviour of the samples that up to a maximal strain of $5\%$ strain despite the non-linearity of the rubber-like base material in the same strain range. This indicates the samples have an expected structural deformation where different parts of the ``\textit{lattice}'' behave as rigid struts and deformable hinges. This effect will be highlighted by the DIC measurements discussed later.\smallskip\\
One can directly observe a lateral expansion during the tensile extension indicating a negative Poisson's ratio for all the samples. The precise measurements of the Poisson's ratio corresponding to a single central unit cell are presented in \Cref{APR}(b). The precise method for the computation of the Poisson's ratio of a single unit cell from DIC measurements was based on periodic homogenisation assumptions and the details are presented in the appendix \ref{appendix}. The results show that the initial effective Poisson's ratio was for all samples close to the announced values in the optimisation process and was not degraded during the manufacturing process. During tensile loading, the effective Poisson's ratio tends to increase, indicating a decrease of the ``auxeticity'' of the samples of up to increases by $10\%$ for a $5\%$ strain.\smallskip\\
Finite element computation were undertaken under the assumption of small strains, large displacements and plane stress using the finite element solver {\tt Cast3M} 2018 (\href{http://www-cast3m.cea.fr}{\texttt{http://www-cast3m.cea.fr}}). The mesh was obtained using image processing from the binarized images of the optimal level set function and completed to the sample geometry. The elastic material behaviour was defined as the tangent behaviour at the origin of the tensile curve of the material. The sample was loaded with a given resultant force at the clamps of the tensile machine.\smallskip\\
Let us first remark, that the evolution is close to predictions of the deformation of the samples obtained by the finite element method under the assumption of large displacements. Second, one can remark that the evolution of the Poisson's ratio with applied strain has already been observed and discussed in \cite{Clausen2015} on polymeric filament structure. Moreover, they arrived to correct the phenomenon up to $20 \%$ strain using a nonlinear material behaviour in the optimisation process, see \cite{Clausen2015,Wang2014} for more details on the subject. In the case of the optimisation procedure presented here, the extension to non-linear material behaviour is currently under investigation and will appear in subsequent work of the authors.
\begin{figure}
\centering
\subf{\includegraphics[width=0.8\textwidth]{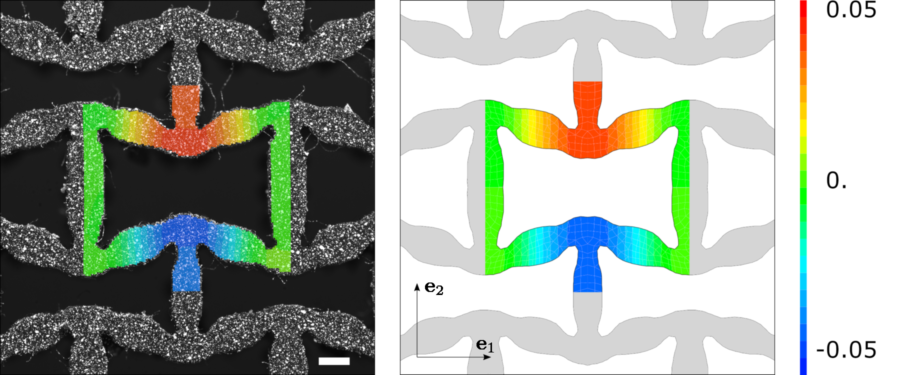}}{(a)}
\subf{\includegraphics[width=0.8\textwidth]{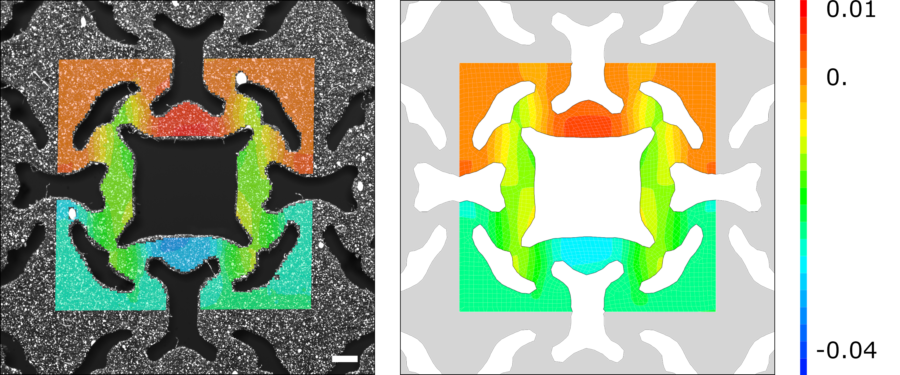}}{(b)}
\subf{\includegraphics[width=0.8\textwidth]{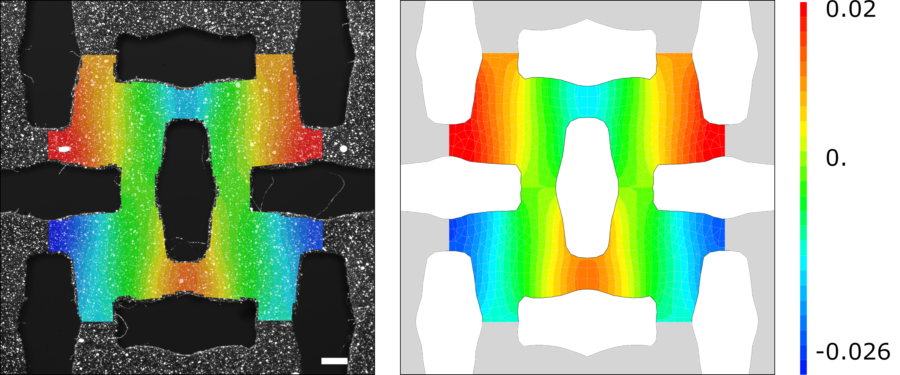}}{(c)}
\caption{Plots of the dimensionless values of the transverse displacement field along $e_2$ for each unit cell of the optimised structure. The samples are loaded along $e_1$ at $5\%$ effective strain. The displacement field for the images on the left were measured using Digital Image Correlation, while for the images on the right using finite element analysis. Image (a) is the optimised structure of Example 1, image (b) the optimised structure of Example 2, image (c) is the optimised structure of Example 3. Scale bar in all images is $1 \, \mathrm{mm}$.}
\label{fig:disp}
\end{figure}
\begin{table}[b]
\begin{tabularx}{\textwidth}{*{1}{>{\centering \arraybackslash}X }|
                             *{6}{>{\centering \arraybackslash}X }}
$\vec{\varepsilon}^H_h$ & DIC$^1$ [\%] & FEM$^1$ [\%] & DIC$^2$ [\%] & FEM$^2$ [\%] & DIC$^3$ [\%] & FEM$^3$ [\%] \\[3pt]
\hline
\vspace{3pt}
$\displaystyle
	\begin{matrix}
    	\mean{\varepsilon_{11}}_h \\[3pt]
    	\mean{\varepsilon_{11}}_h \\[3pt]
   		\mean{\varepsilon_{12}}_h \\
  	\end{matrix}
$
&
\vspace{3pt}
$\displaystyle
	\begin{matrix}
    \phantom{-}1.250 \\[3pt]
   	         - 1.010 \\[3pt]
   	\phantom{-}0.039 \\
  	\end{matrix}
$
&
\vspace{3pt}
$\displaystyle
	\begin{matrix}
    \phantom{-}0.543 \\[3pt]
   		     - 0.284 \\[3pt]
   	\phantom{-}0.004 \\
  	\end{matrix}
$
&
\vspace{3pt}
$\displaystyle
	\begin{matrix}
    \phantom{-}0.526 \\[3pt]
   		     - 0.450 \\[3pt]
   	\phantom{-}0.053 \\
  	\end{matrix}
$
&
\vspace{3pt}
$\displaystyle
	\begin{matrix}
    \phantom{-}0.404 \\[3pt]
   	         - 0.204 \\[3pt]
   	\phantom{-}0.002 \\
  	\end{matrix}
$
&
\vspace{3pt}
$\displaystyle
	\begin{matrix}
    \phantom{-}0.937 \\[3pt]
   	         - 0.643 \\[3pt]
   	\phantom{-}0.010 \\
  	\end{matrix}
$
&
\vspace{3pt}
$\displaystyle
	\begin{matrix}
    \phantom{-}0.828 \\[3pt]
   	         - 0.407 \\[3pt]
   	\phantom{-}0.001 \\
  	\end{matrix}
$
\end{tabularx}
\caption{Mean strain components of the base material measured by DIC and computed by FEA on the unit cells from \Cref{fig:disp} for the structures in Examples 1-3, denoted by the superscript. The measurement corresponds to the maximal loading with an effective strain of 5\%.}
\label{table:epsilon:mean}
\end{table}
The displacements fields obtained using DIC permit a further comparison with predictions and give an insight of the deformation mechanism of the samples, \textit{i.e.} how the structure moves and deforms. \Cref{fig:disp} displays the measured and the computed vertical displacement, \textit{i.e.} the $u_y$ displacement component of the central unit-cell. A comparison of the values and the shapes of the colour maps exhibits a good match between the measurements and finite element prediction. Moreover the displacement fields permit to better understand the local movements of the micro-structure which conducts to the global auxeticity effect by combining almost rigid regions submitted to translations and rotations with local concentrated deformation exhibiting local hinges. A further comparison in terms on mean displacements over a unit cell between DIC measurements and FEM computations at the maximal loading of $5\%$ strain is given in \Cref{table:epsilon:mean}. The mean was computed only over the base material of a unit cell and does represent the mean deformation of the later. Nevertheless, this mean value represents the mean loading of the base material and one can notice that the micro-structure does not leave the region of $1\%$ strain of the uniaxial tensile response of the base material displayed in \Cref{GM08B}. In \Cref{fig:disp} one can equally notice the excellent quality of the printing process as the edges of the printed shapes observed on left column of \Cref{fig:disp} are close to the edges of the shape represented by the level set and displayed here as the border of the finite element meshes on the right column of \Cref{fig:disp}.\smallskip\\
By looking at the strain field of a unit cell (see \Cref{fig:strain}), computed by finite elements from the displacement field in both full-field measurement and simulation, we can notice that the strain field is mostly concentrated on the hinges of the structure. This further emphasises the predominance of structural deformation as a lattice structure with rods and hinges at small strain. A subtle effect of this prevalence is the small effect of the out of plane strain, which should otherwise be perceived as a difference between the 2D modelling during the optimisation process and the complete 3D character of the polymer sheets. Let us also recall that the polymer is practically incompressible and exhibit therefore an important variation of thickness under tensile loading.\smallskip
\\
The next steps in the analysis of the micro-architectured material is the complete experimental measurement of its elastic tensor. Let us recall that the effective constitutive law \eqref{eq:sig:A:eps} or alternatively \eqref{eq:sig:A:eps:coord} is a linear relation between the components of the effective stress and strain, from which the elastic moduli could be identified by a least square fitting. The main difficulty is that only the effective strain, $\vec{\varepsilon}^H$, can be directly measured from the experiment, see values in \Cref{table:meaneps}. However, as suggested in \cite{Rethore2017}, the effective stress, $\vec{\sigma}^H$, can be numerically computed from the experimental applied forces if the geometry and the constitutive behaviour of the base material. As a consequence, $\mathbb{C}^H$, the effective elastic tensor of the design phase is obtained as a linear fit from $\vec{\varepsilon}^H$ and $\vec{\sigma}^H$. The computation can be performed on several unit cells of the specimen. In order to compare the values of the elasticity tensor $\mathbb{C}^H$ computed in the design phase, the resultant forces were normalised.
\begin{table}[b]
\centering
\begin{tabularx}{\textwidth}{c | *{5}{>{\centering \arraybackslash} X}}
Example & $\mathbb{C}^H(\omega)$ & $\mathbb{C}^{H,exp}(\omega)$ & $\nu^*$ & $\nu^{*,exp}$ & Shape $\omega$\\
\hline
2 &
{$\displaystyle
	\begin{pmatrix}
    	\phantom{-}0.12 & -0.05 & 0 \\
    	-0.05 & \phantom{-}0.12 & 0 \\
   		0 & 0 & G \\
  	\end{pmatrix}$} &
{$\displaystyle
	\begin{pmatrix}
    	\phantom{-}0.10 & -0.044 & 0 \\
   		-0.044 & \phantom{-}0.10 & 0 \\
   		0 & 0 & G \\
  	\end{pmatrix}$} &
-0.42 &
-0.44 &
\vspace{0.1cm}
\includegraphics[width=0.1\columnwidth]{ex2_micro.pdf}\\
3 &
{$\displaystyle
	\begin{pmatrix}
    	\phantom{-}0.19 & -0.09 & 0 \\
    	-0.09 & \phantom{-}0.19 & 0 \\
   	0 & 0 & G \\
  	\end{pmatrix}$} &
{$\displaystyle
	\begin{pmatrix}
    	\phantom{-}0.204 & -0.09 & 0 \\
    	-0.09 & \phantom{-}0.204 & 0 \\
   		0 & 0 & G \\
  	\end{pmatrix}$} &
-0.47 &
-0.44 &
\vspace{0.1cm}
\includegraphics[width=0.1\columnwidth]{ex3_micro.pdf}
\end{tabularx}
\caption{Comparison between the effective $\mathbb{C}^H(\omega)$ (see also \Cref{table:Hcoef}) and measured elasticity tensor $\mathbb{C}^{H,exp}(\omega)$ displayed in the left and center column respectively. The right column displays the optimal shape in each case. We recall that the measured elasticity tensor $\mathbb{C}^{H,exp}(\omega)$ was determined by combining DIC measurements and FEM computations.}
\label{table:meaneps}
\end{table}\smallskip\\
Specimens in examples 2 and 3 have a cubic material symmetry, which leads to a system of three equations with three unknowns for each unit cell. One has to identify three moduli $C_{1111} = C_{2222}, C_{1122}, C_{1212}$ using the ${11}, {22}, {12}$ strain and stress components. The estimated elastic moduli on the central unit cell, \textit{i.e.} with position (3,3) and coloured red in \Cref{fig:3D}, are displayed in \Cref{table:meaneps}. The $C_{1212}$ moduli is missing as the signal to noise ratio of the effective shear strains and stresses was to small to provide meaningful value.

\section{Conclusion}
In this work we used a topology optimisation method to design optimal shapes that achieve a negative Poisson's ratio. By removing certain material constraints, \textit{e.g.} isotropy, from the algorithm we expanded the space of admissible shapes and as a result the algorithm was able to attain shapes with a Poisson's ratio below -1. The effective elasticity tensor characterising the material with Poisson's ratio below -1 is orthotropic and although the theoretical problem of reachable elasticity tensors has been solved in the seminal work of Milton \& Cherkaev \cite{Milton1995} the algorithm suggests that the more we expand the space of admissible shapes by allowing shapes to deviate from isotropic symmetry the closer to the \textit{stability} bounds the effective material approaches (see \Cref{fig:Stability}).\smallskip\\
The results showed that optimal shapes could be directly printed without additional enhancement of the surface, which is a direct consequence of the smoothed interface technique used in the optimisation. Moreover the manufactured materials had the designed mechanical behaviour. The targeted elastic moduli and the underlying Poisson's ratios have been experimentally attained and the local material behaviour was close to predictions.\smallskip\\
The local displacement field computed and measured on the micro-structure showed that measurements match numerical predictions. Moreover one can observe from the strain field that the global deformation is composed of rigid regions an localised hinges, indicating that the structures behaves as rotating rigid units.

\begin{acknowledgements}
This work is financed by the french-swiss ANR-SNF project MechNanoTruss (ANR-15-CE29-0024-01). The authors would like to express their gratitude to Chiara Daraio for fruitful discussion on the design of lattice structures and to Grégoire Allaire and Georgios Michailidis for lending their expertise on the numerical and algorithmic issues of the optimisation.\bigskip\\
\textbf{Conflict of interest} The authors declare that they have no conflict of interest.
\end{acknowledgements}

\vfill

\pagebreak
\appendix

\section{Computation of the effective Poisson's ratio} \label{appendix}
This appendix reviews the mathematical approach that was used to measure/compute the effective Poisson's ratio of a unit cell in both measurement by Digital Image Correlation and numerical estimation using a finite element method. For the following computation, we place ourselves in the case of small strain assumption.
\begin{figure}[h]
\centering
\includegraphics[width=0.4\textwidth]{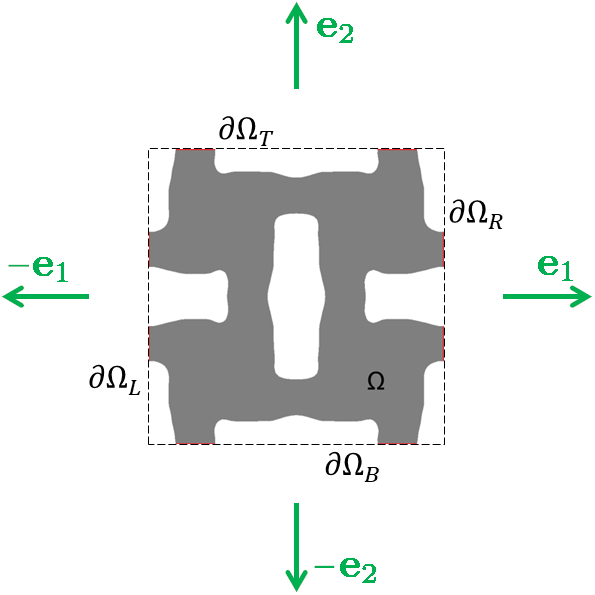}
\caption{Representation of a unit cell.}
\label{APR_computation}
\end{figure}\\
The effective material is supposed to carry a natural orthotropic material behaviour. The effective Poisson's ratio $\nu_{12}$, characterising the transverse strain of the structure in the direction $(O,\vec{e_2})$ axis when stretched in the direction $(O,\vec{e_1})$, is defined as:
\begin{equation}
\nu_{12}^*=\frac{C^H_{1122}}{C^H_{2222}}
\label{eq:APR_def}
\end{equation}
We remind that $C^H_{1122}$ and $C^H_{2222}$ are coefficients of the effective elastic stiffness tensor. In general $\nu_{12} \neq \nu_{21}$. During a uniaxial tensile test in the direction $(O,\vec{e_1})$, equation \eqref{eq:APR_def} yields to the negative of the ratio of macroscopic transverse strain to macroscopic axial strain:
\begin{equation}
\nu_{12}^* = - \frac{\varepsilon^H_{22}}{\varepsilon^H_{11}}
\end{equation}
In the small strain assumption, the strain field can be linearised as: 
\begin{equation}
\ten[2]{\varepsilon}^H = \mean{\ten[2]{\varepsilon}}_{\Omega} = \frac{1}{2} \left( \mean{\ten[2]{F}}_{\Omega}^T + \mean{\ten[2]{F}}_{\Omega} \right) - \ten[2]{I}
\label{eq:eps_ss}
\end{equation}
where $\ten[2]{F}$ is the average transformation gradient. Considering the small strain assumption:
\begin{equation}
\mean{\ten[2]{F}}_{\Omega} = \frac{1}{V_{\Omega}} \int_{\Omega} (\ten[2]{I} +\vec{\nabla} \vec{u}) d \Omega
\end{equation}
Using Ostrogradsky's theorem, we can express the transformation gradient at the boundary $\partial \Omega$:
\begin{equation}
\mean{\ten[2]{F}}_{\Omega} = \frac{1}{V_{\Omega}} \left( \int_{\Omega} \ten[2]{I} \, d\Omega + \oint_{\Gamma} \vec{u} \otimes \vec{n} \, d\Gamma \right)
\end{equation}
Study of a unit cell.
\begin{equation}
\mean{\ten[2]{F}}_{\Omega} = \ten[2]{I} + \frac{1}{V_{\Omega}} \left(
\int_{\partial\Omega_T} \vec{u} \otimes   \vec{e_2} \, d\Gamma +
\int_{\partial\Omega_B} \vec{u} \otimes (-\vec{e_2})\, d\Gamma +
\int_{\partial\Omega_R} \vec{u} \otimes   \vec{e_1} \, d\Gamma +
\int_{\partial\Omega_L} \vec{u} \otimes (-\vec{e_1})\, d\Gamma \right)
\end{equation}
\begin{equation}
\mean{\ten[2]{F}}_{\Omega} =  \ten[2]{I} + \frac{1}{V_{\Omega}}
\begin{bmatrix}
\displaystyle \int_{\partial\Omega_R} u_1\, d\Gamma - \int_{\partial\Omega_L} u_1\, d\Gamma & 
\displaystyle \int_{\partial\Omega_T} u_1\, d\Gamma - \int_{\partial\Omega_B} u_1\, d\Gamma 
\\[12pt]
\displaystyle \int_{\partial\Omega_R} u_2\, d\Gamma - \int_{\partial\Omega_L} u_2\, d\Gamma & 
\displaystyle \int_{\partial\Omega_T} u_2\, d\Gamma - \int_{\partial\Omega_B} u_2\, d\Gamma
\end{bmatrix}
\end{equation}
Thus from equation \eqref{eq:eps_ss}: 
\begin{equation}
\begin{cases}
\displaystyle \varepsilon_{11} = \frac{1}{V_{\Omega}} 
\left(\int_{\partial\Omega_R} u_1\, d\Gamma - \int_{\partial\Omega_L} u_1\, d\Gamma \right)
\\[12pt]
\displaystyle \varepsilon_{22} = \frac{1}{V_{\Omega}}
\left(\int_{\partial\Omega_T} u_2\, d\Gamma - \int_{\partial\Omega_B} u_2\, d\Gamma \right)
\end{cases}
\end{equation}
For each edge of the square unit cell, the integral of the contour is computed by integrating the displacement of the material in contact with the edge. In other words, the void phase is not considered in the computation.
\begin{equation}
\nu_{12}^* = -\frac{\displaystyle \int_{\partial\Omega_T} u_2\, d\Gamma -
                                  \int_{\partial\Omega_B} u_2\, d\Gamma}
                   {\displaystyle \int_{\partial\Omega_R} u_1\, d\Gamma -
                                  \int_{\partial\Omega_L} u_1\, d\Gamma}
\label{eq:APR_formula}
\end{equation}
In practice, using a finite element method, equation \eqref{eq:APR_formula} becomes : 
\begin{equation}
\nu_{12}^* = -\frac{ \displaystyle \frac{1}{N_T} \sum \limits_{i = 1}^{N_T} u_2^i - \frac{1}{N_B} \sum \limits_{b}^{N_B} u_2^i}{ \displaystyle \frac{1}{N_R} \sum \limits_{i = 1}^{N_R} u_1^i - \frac{1}{N_L} \sum \limits_{i = 1}^{N_L} u_1^i}
\end{equation}
where $N_i, i \in \{ T,B,R,L\}$ are respectively the number of nodes on top, bottom, right and left edges.
\end{document}